\documentclass[11pt,a4paper]{article}
\pdfoutput=1
\usepackage{jcappub}
\usepackage[T1]{fontenc}
\usepackage{graphicx}
\usepackage{bm}
\usepackage{amsmath,amsfonts,amssymb}
\usepackage{mathtools}
\usepackage{braket}
\usepackage{setspace} 
\usepackage{here}
\usepackage{xcolor}

\usepackage{soul}
\usepackage[normalem]{ulem}
\usepackage{cancel}

\begin{document}


\title{Affleck-Dine leptogenesis scenario for resonant production of sterile neutrino dark matter}
\author[a]{Kentaro Kasai,}
\author[a,b]{Masahiro Kawasaki,}
\author[c]{and Kai Murai}
\affiliation[a]{ICRR, University of Tokyo, Kashiwa, 277-8582, Japan}
\affiliation[b]{Kavli IPMU (WPI), UTIAS, University of Tokyo, Kashiwa, 277-8583, Japan}
\affiliation[c]{Department of Physics, Tohoku University, Sendai, 980-8578, Japan}

\abstract{Sterile neutrino is a fascinating candidate for dark matter. 
In this paper, we examine the Affleck-Dine (AD) leptogenesis scenario generating a large lepton asymmetry,
which can induce
the resonant production of sterile neutrino dark matter
via the Shi-Fuller (SF) mechanism.
We also revisit the numerical calculation of the SF mechanism and the constraints from current X-ray and Lyman-$\alpha$ forest observations.
We find that the AD leptogenesis scenario can explain the production of sterile neutrino dark matter by incorporating a non-topological soliton with a lepton charge called L-ball.
Finally, we discuss 
an enhancement of second-order gravitational waves at the L-ball decay and investigate the testability of our scenario with future gravitational wave observations.
}


\emailAdd{kkasai@icrr.u-tokyo.ac.jp}
\emailAdd{kawasaki@icrr.u-tokyo.ac.jp}
\emailAdd{kai.murai.e2@tohoku.ac.jp}

\begin{flushright}
    TU-1222
\end{flushright}

\maketitle

\section{Introduction}
\label{sec: intro}
Sterile neutrinos with masses of the $\rm{keV}$ scale are one of the attractive candidates for dark matter (for a review, see Refs.~\cite{Boyarsky:2018tvu,Abazajian:2017tcc}).
Sterile neutrinos can be produced via mixing with the standard model (SM) neutrinos (``active'' neutrinos)~\cite{Dodelson:1993je, Abazajian:2001nj,Abazajian:2001vt,Shi:1998km}.
The simplest and most well-known mechanism to produce sterile neutrino dark matter is the Dodelson-Widrow (DW) mechanism~\cite{Dodelson:1993je, Abazajian:2005gj}.
In this mechanism, sterile neutrinos are produced through neutrino oscillations between active and sterile neutrinos in the early universe. 
However, this scenario is excluded by X-ray observations of the radiative decay of sterile neutrinos~\cite{Dolgov:2000ew,Abazajian:2001vt,Boyarsky:2005us,Boyarsky:2006fg,Shrock:1974nd,denHerder:2009sxr} and by Lyman-$\alpha$ constraints from structure formation~\cite{Baur:2017stq,Viel:2005qj}.

An alternative scenario is the resonant production of sterile neutrinos called the Shi-Fuller (SF) mechanism~\cite{Shi:1998km}. 
This mechanism assumes a non-zero lepton asymmetry in the active neutrino sector.
Then, the effective mixing angle between active and sterile states 
is resonantly enhanced at a certain cosmic temperature due to the lepton asymmetry.
Consequently, we can explain all dark matter with a smaller vacuum mixing angle between active and sterile states than that required in the DW mechanism~\cite{Gelmini:2019clw,Abazajian:2001nj,Abazajian:2001vt,Boyarsky:2009ix,Kishimoto:2008ic,Laine:2008pg, Ghiglieri:2015jua}.
Thus, we expect that this scenario evades the constraints from X-ray observations. 
Importantly, this scenario requires a lepton-to-entropy ratio $\gtrsim \mathcal{O}(10^{-5})$ much larger than the observed baryon-to-entropy ratio ($\sim 10^{-10}$) to account for all dark matter through the resonance effect.

The possibility of such a large lepton asymmetry is also well motivated by the recent observation of helium-4 in metal-poor galaxies, which suggests the existence of a large lepton asymmetry before big bang nucleosynthesis (BBN)~\cite{Matsumoto:2022tlr}.
However, 
if the lepton asymmetry is generated when the sphaleron process is in chemical equilibrium, baryon and lepton asymmetries must be of the same order. 
Therefore, we must consider the scenario in which the large lepton asymmetry
is generated after the electroweak phase transition. 

In this paper, we consider the Affleck-Dine (AD) leptogenesis scenario for the production of a large lepton asymmetry.
In this scenario, the lepton charge is confined in non-topological solitons called L-balls and is protected against the sphaleron process.
Afterward, L-balls decay and release the lepton charge to the thermal plasma after the electroweak phase transition~\cite{Kawasaki:2002hq,Kawasaki:2022hvx}.
This scenario can account for a lepton-to-entropy ratio as large as $\mathcal{O}(10^{-3})$ in contrast to leptogenesis through the decay of heavy sterile neutrinos, which typically explains a lepton-to-entropy ratio of at most $\sim 7\times 10^{-4}$~\cite{Shaposhnikov:2008pf}.  

Interestingly, our scenario has another observational implication.
To produce a large lepton asymmetry the L-balls dominate the energy density in the universe before the decay.
Since a sudden transition from a matter-dominated era 
to a radiation-dominated era occurs at L-ball decay, 
gravitational waves generated from the second-order effect of the scalar mode are enhanced on small scales. 
We investigate the testability of our scenario by future observations of gravitational waves through $\mu$Ares~\cite{Sesana:2019vho}, and THEIA~\cite{Theia:2019non}.

In this paper, we also revisit the numerical calculation of the SF mechanism to discuss the observational constraint. 
We numerically evaluate the required vacuum mixing angle between the active and sterile states for a given sterile neutrino mass $m_s$ and the initial lepton asymmetry in the active neutrino sector to explain all dark matter.
In this calculation, we take into account the time dependence of the relativistic degree of freedom, which is intrinsically important because the resonance occurs around the QCD phase transition for the parameters of interest.
We find that the lepton-to-entropy ratio is required to be $\gtrsim \mathcal{O}(10^{-4})$ to evade both the X-ray and Lyman-$\alpha$ constraints.

The rest of this paper is organized as follows. 
In Sec.~\ref{sec: calculation of Shi Fuller mechanism}, we review the production of sterile neutrinos through the SF mechanism and show the numerical results.
We explain the leptogenesis in the AD mechanism in Sec.~\ref{sec: AD leptogenesis}.
We discuss the observational implications of our scenario in Sec.~\ref{sec: observational implications}.
Sec.~\ref{sec: conclusion} is devoted to the conclusion of our results.

\section{Numerical study of Shi-Fuller mechanism revisited}
\label{sec: calculation of Shi Fuller mechanism}

In this section, we revisit the numerical calculation of sterile neutrino production by the SF mechanism, which was originally proposed in Ref.~\cite{Shi:1998km} (see also Refs.~\cite{Abazajian:2001nj,Abazajian:2004aj,Boyarsky:2009ix,Gelmini:2019clw,Ghiglieri:2015jua,Laine:2008pg,Kishimoto:2008ic}).

\subsection{Shi-Fuller mechanism}

First, we derive the basic equations of the SF mechanism following the discussion in Ref.~\cite{Dolgov:2002wy}.
Here, we consider the mixing between one generation of sterile neutrinos and one flavor of active neutrinos for simplicity.
Then, the sterile neutrino state mixes with the active neutrino state as 
\begin{equation}
\begin{aligned}
    \ket{\nu_a}&=\cos\theta\ket{\nu_1}+\sin\theta\ket{\nu_2}
    \ , 
    \\
    \ket{\nu_s}&=-\sin\theta\ket{\nu_1}+\cos\theta\ket{\nu_2}
    \ ,
\end{aligned}
\end{equation}
where $\ket{\nu_a}$ and $\ket{\nu_s}$ denote the active and sterile neutrino states, respectively, and $\ket{\nu_1}$ and $\ket{\nu_2}$ denote the two mass eigenstates.

The effective description of the kinetic equation of neutrino states is given as~\cite{Dolgov:2002wy,Shaposhnikov:2008pf}
\begin{equation}
    i(\partial_t-Hp\partial_p) \rho
    =
    \left[ \mathcal{H},\rho \right]
    -i\{ \Gamma,\rho-\rho_{\rm{eq}} \}
    \ ,
    \label{eq : density operator equation}
\end{equation}
where $H$ is the Hubble parameter, $p$ is the physical momentum, $\mathcal{H}$ is the Hamiltonian, and $[\cdot\, ,\, \cdot]$ and $\{\cdot\, , \,\cdot\}$ denote commutation and anti-commutation relations, respectively.
Here, $\rho$ is the density operator in the flavor basis, i.e., the active and sterile neutrinos, and its components are written as 
\begin{equation}
    \rho (t, p)
    =
    \begin{pmatrix}
        \rho_{aa} & \rho_{as} \\
        \rho_{as}^* & \rho_{ss} \\
    \end{pmatrix}
    \ .
    \label{eq : definition of density matrix}
\end{equation}
Here, $\rho_{aa}$ and $\rho_{ss}$ correspond to the distribution function of the active and sterile neutrinos, respectively, and $\rho_{as}^*$ is the complex conjugate of $\rho_{as}$.
$\rho_{\rm{eq}}$ is the thermal distribution given by
\begin{equation}
    \rho_{\rm{eq}}
    =
    \begin{pmatrix}
        f_{\rm{eq}}(\sqrt{m_a^2+p^2},\mu_{\nu_a}) & 0 \\
        0 & f_{\rm{eq}}(\sqrt{m_s^2+p^2},\mu_{\nu_s}) \\
    \end{pmatrix}
    \ ,
    \label{eq : density matrix describing thermal distribution}
\end{equation}
where $m_a$ and $m_s$ are the active and sterile neutrino masses, respectively, $f_{\rm{eq}}(E,\mu) \equiv 1/(\exp[(E-\mu)/T] + 1)$ is the Fermi-Dirac distribution function, and $\mu_{\nu_a}$ and $\mu_{\nu_s}$ are the chemical potentials of the active and sterile neutrinos, respectively. 
In the following, we approximate that $\sqrt{m_a^2+p^2}\simeq\sqrt{m_s^2+p^2}\simeq p$ since both the active and sterile neutrinos are relativistic at the epoch of interest.
$\Gamma$ is a Hermitian matrix associated with the scattering rate of neutrinos in the thermal plasma.
Since sterile neutrinos do not interact with the thermal plasma, $\Gamma$ has a form written as
\begin{equation}
    \Gamma
    =
    \frac{1}{2}
    \begin{pmatrix}
        \Gamma_{\nu_a} & 0 \\
               0 & 0 \\
    \end{pmatrix}
    \ ,
    \label{eq : matrix describing scattering}
\end{equation}
where $\Gamma_{\nu_a}$ is the total thermal width of the active neutrino given by
\begin{align}
    \Gamma_{\nu_a} (p,T)
    &=
    y_a(T) G_{\rm{F}}^2 p T^4 
    \ .
    \label{eq : thermal width}
\end{align}
Here, $G_{\rm{F}}$ is the Fermi coupling constant, and $y_a(T)$ is a factor depending on the degrees of freedom in the thermal plasma.
In the flavor basis, the Hamiltonian $\mathcal{H}$ is written as
\begin{equation}
    \mathcal{H}
    =
    \begin{pmatrix}
        V_a-\frac{m_s^2}{4p}\cos2\theta & \frac{m_s^2}{4p}\sin2\theta 
        \vspace{5pt} \\
        \frac{m_s^2}{4p}\sin2\theta & \frac{m_s^2}{4p}\cos2\theta
        \end{pmatrix}
        \ ,
    \label{eq : Hamiltonian matrix}
\end{equation}
where $V_a$ denotes the effective potential due to the finite temperature and density effects on the self-energy of the active neutrino.
The first term in the right-hand side of Eq.~\eqref{eq : density operator equation} describes the neutrino oscillation effect, and the second term describes the scattering process of active neutrinos.

Consequently, Eq.~\eqref{eq : density operator equation} can be rewritten as
\begin{align}
    i(\partial_t-Hp\partial_p)\rho_{aa}
    &=
    \frac{m_s^2}{4p}\sin2\theta(\rho_{as}^*-\rho_{as})-i\Gamma_{\nu_a}(\rho_{aa}-f_{\rm{eq}})
    \ , 
    \label{eq : density operator equation of rho aa}
    \\
    i(\partial_t-Hp\partial_p)\rho_{ss} 
    &=
    \frac{m_s^2}{4p}\sin2\theta(\rho_{as}-\rho_{as}^*)
    \ , 
    \label{eq : first line in density operator equation}
    \\
    i(\partial_t-Hp\partial_p)\rho_{as} 
    &=
    \frac{m_s^2}{4p}\sin2\theta(\rho_{ss}-\rho_{aa})
    +
    \left(
        V_a -\frac{m_s^2}{2p}\cos2\theta
        -i\frac{\Gamma_{\nu_a}}{2}
    \right) \rho_{as}
    \ .
    \label{eq : second line in density operator equation}
    \end{align}
Since $\Gamma_{\nu_a}$ is much larger than $H$ and $m_s^2/p$ for the temperatures of interest, Eq.~\eqref{eq : density operator equation of rho aa} leads to $\rho_{aa} \simeq f_\mathrm{eq}(p,\mu_{\nu_a})$.
Furthermore, we assume that the left-hand side of Eq.~\eqref{eq : second line in density operator equation} vanishes. 
This is called stationary point approximation~\cite{Dolgov:2002wy}, and we will discuss the validity of this approximation in Sec.~\Ref{subsec : Validity of stationary point approximation}.
Under the stationary point approximation, we obtain 
\begin{align}
    (\partial_t-Hp\partial_p)\rho_{ss}
    &=
    \frac{1}{4}
    \Gamma_{\nu_a}\sin^22\theta
    \left[ 
        \left( \cos2\theta-\frac{2p}{m_s^2}V_a \right)^2
        +\frac{p^2}{m_s^4}\Gamma_{\nu_a}^2
    \right]^{-1}
    (\rho_{aa}-\rho_{ss})
    \, .
    \label{eq : time evolution of rho ss}
\end{align}
Similarly, the distribution function of the anti-sterile neutrino follows
\begin{align}
    (\partial_t-Hp\partial_p)\rho_{\bar{s}\bar{s}}
    &=
    \frac{1}{4}
    \Gamma_{\nu_a} \sin^2 2\theta
    \left[
        \left( \cos2\theta-\frac{2p}{m_s^2}V_{\bar{a}} \right)^2
        + \frac{p^2}{m_s^4}\Gamma_{\nu_a}^2
    \right]^{-1}
    (\rho_{{\bar{a}}{\bar{a}}} - \rho_{\bar{s}\bar{s}})
    \ ,
    \label{eq : time evolution of rho bar ss}
\end{align}
where $V_{\bar{a}}$ is the effective potential of the anti-active neutrino. 
Here, we approximated
the scattering rate of the anti-active neutrino by
$\Gamma_{\nu_a}$ since the contribution from the chemical potential $\mu_{\nu_a}$ is negligible. 

To solve the Boltzmann equation, it is convenient to define a comoving momentum $\propto pa$.
From the conservation of the total entropy density, we define a dimensionless comoving momentum by
\begin{equation}
    \epsilon
    \equiv
    \left( \frac{g_{*,s}(T_{i})}{g_{*,s}(T)} \right)^{1/3}
    \frac{p}{T}
    \ ,
    \label{eq : dimensionless mode}
\end{equation}
where $g_{*,s}$ is the effective number of relativistic degrees of freedom for entropy, $T$ is cosmic temperature, and $T_i$ is the reference temperature.
Note that $\epsilon$ is generally different from $\epsilon_\mathrm{phys} \equiv p/T$ due to the time evolution of $g_{*,s}$.

Defining $f_{\nu_a}\equiv \rho_{aa}$, $f_{{\bar{\nu}}_a}\equiv \rho_{\bar{a}\bar{a}}$, and $f_{\nu_s}\equiv \rho_{ss}+\rho_{\bar{s}\bar{s}}$, and using Eqs.~\eqref{eq : time evolution of rho ss} and \eqref{eq : time evolution of rho bar ss}, 
we obtain the production rate of sterile neutrinos as
\begin{align}
    \frac{{\rm{d}}}{{\rm{d}}t} f_{\nu_s}(\epsilon,T)
    =
   \Gamma_{\nu_a}(p,T)
    \left[
        \theta_M^2(\epsilon,T) f_{\nu_a}(\epsilon,T)
        +
        \bar{\theta}_M^2(\epsilon,T) f_{\bar{\nu}_a}(\epsilon,T)
    \right]
    \ ,
    \label{master equation of nus distribution}
\end{align}
where $\theta_M$ and $\bar{\theta}_M$ are effective mixing angle in the thermal plasma defined by
\begin{equation}
\begin{aligned}
   \theta_M^2(\epsilon,T)
   &\equiv
    \theta^2
    \left[
        \left( 1 - \frac{2p}{m_s^2}V_{a}(\epsilon,T) \right)^2
        + \frac{p^2\Gamma_{\nu_a}^2(\epsilon,T)}{m_s^4}
    \right]^{-1}
    \ ,
   \\
    \bar{\theta}_M^2(\epsilon,T)
   &\equiv
    \theta^2
    \left[
        \left(1 - \frac{2p}{m_s^2}V_{\bar{a}}(\epsilon,T) \right)^2
        + \frac{p^2\Gamma_{\nu_a}^2(\epsilon,T)}{m_s^4}
    \right]^{-1}
    \ .
    \label{definition of thetaM}
\end{aligned}
\end{equation}
Here, we used the approximations of $\cos\theta\simeq 1$ and $\sin\theta\simeq \theta$, and used $f_{\nu_a}, f_{\bar{\nu}_a}\gg f_{\nu_s}$, which holds for the parameter region of our interest. 
The thermal self-energies of the active and anti-active neutrinos, $V_a$ and $V_{\bar{a}}$, are respectively given by~\cite{Notzold:1987ik}
\begin{equation}
\begin{aligned}
    V_a
    &= 
    \sqrt{2}G_{\rm{F}} \Bigl(2(n_{\nu_a}-n_{\bar{\nu}_a})
    + \sum_{b \neq a}(n_{\nu_b}-n_{\bar{\nu}_b})\Bigr)-B_ap T^4
    \\
    &= 
    \sqrt{2}G_{\rm{F}} \Bigl(2L_{\nu_a}
    + \sum_{b \neq a}L_{\nu_{b}}\Bigr)s_{\rm{tot}}
    -
    B_a\epsilon_{\rm{phys}}(T)T^5,
    \\
    V_{\bar{a}} 
    &=
    \sqrt{2}G_{\rm{F}} \Bigl(-2(n_{\nu_a}-n_{\bar{\nu}_a})
    + \sum_{b \neq a}(-n_{\nu_b}+n_{\bar{\nu}_b})\Bigr)-B_ap T^4
    \\
    &= 
    -\sqrt{2}G_{\rm{F}} \Bigl(2L_{\nu_a}
    + \sum_{b \neq a}L_{\nu_{b}}\Bigr)s_{\rm{tot}}
    -
    B_a\epsilon_{\rm{phys}}(T)T^5,
    \label{eq : finite density potential}
\end{aligned}
\end{equation}
where
$n_{\nu_a}$ and $n_{\bar{\nu}_a}$ are the number densities of 
$\nu_a$ and $\bar{\nu}_a$, respectively, and $s_{\rm{tot}}$ is the total entropy density.
Here, the subscripts $a,b\, (=e,\mu,\tau)$ denote the flavors of the active neutrinos, and $B_a$ is a constant taking a value of $B_e=10.88 \times 10^{-9}\,{\rm{GeV}}^{-4}$ for $a=e$~\cite{Gelmini:2019wfp}.
We defined a lepton asymmetry parameter $L_{\nu_a}\equiv (n_{\nu_a}-n_{\bar{\nu}_a})/s_{\rm{tot}}$.
We neglected the contributions of the baryon and charged lepton sectors because they are much smaller than that of the active neutrino sector as discussed in Refs.~\cite{Laine:2008pg,Ghiglieri:2015jua}.

\begin{figure}[t]
\centering
\includegraphics[width=0.75\linewidth]{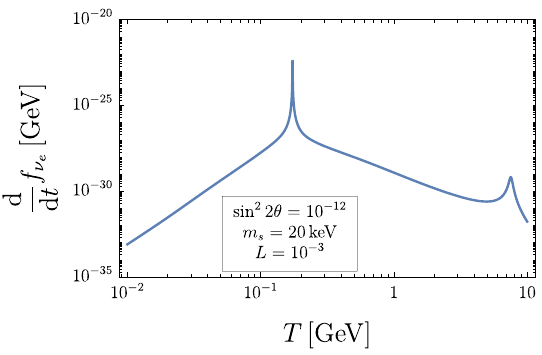}
    \caption{%
    Production rate of sterile neutrinos ${\rm{d}}f_{\nu_s}/{\rm{d}}t$ for $\epsilon=1$ with $T_i=1$\,GeV, $\sin^2 2\theta = 10^{-12}$, $m_s=20$\,keV, and $L=10^{-3}$.
    We see that there are two resonance points at $T\sim0.2$\,GeV and $T\sim7$\,GeV.
    }
    \label{fig : production rate}
\end{figure}

If there is no lepton asymmetry, $V_a$ and $V_{\bar{a}}$ are equal and negative ($V_a = V_{\bar{a}} < 0$), and hence $\nu_s$ and $\nu_{\bar{s}}$ are produced with the same amount via the DW mechanism.
On the other hand, if $L\equiv\,2L_{\nu_a} + \sum_{b \neq a}L_{\nu_{b}} > 0$ is large enough, $V_a$ becomes positive and $\theta_M$ is enhanced at some temperature.
Consequently, $\nu_s$ is resonantly produced while the production of $\nu_{\bar{s}}$ is not so different from the DW mechanism.
If $L < 0$, $\nu_{\bar{s}}$ can be resonantly produced in a similar way.
This is how sterile neutrinos are produced in the SF mechanism.
We show ${\rm{d}}f_{\nu_s}/{\rm{d}}t$ for $\epsilon = 1$ and a fixed value of $L$ in Fig.~\Ref{fig : production rate}. Here, we assumed that sterile neutrinos have a mixing only with electron neutrinos.
We see that there are two resonance points. 
As we will discuss in Sec.~\ref{subsec : analytical study}, the resonance at the lower temperature is generally more important when the resonance effect is dominant in production.
Note that the production rate ${\rm{d}}f_{\nu_s}/{\rm{d}}t$ in the realistic situation is different from that shown in Fig.~\Ref{fig : production rate} because the lepton asymmetry evolves in time.

\subsection{Numerical setup}

In this section, we explain the physical assumptions and the setup in our calculation of the SF mechanism.
In numerical calculations, we focus on the mixing between electron neutrinos and sterile neutrinos.

First of all, we assume that active neutrinos of all flavors $a=e,\mu,\tau$ are in thermal equilibrium and have the Fermi-Dirac distribution given by
\begin{equation}
\begin{aligned}
    f_{\nu_a}(\epsilon,T)
    &=
    \left[
        1+\exp\left(\epsilon_{\rm{phys}}(T)-\xi_{\nu_a}(T)\right)
    \right]^{-1}
    \ , 
    \\
    f_{\bar{\nu}_a}(\epsilon,T)
    &=
    \left[
        1+\exp\left(\epsilon_{\rm{phys}}(T)+\xi_{\nu_a}(T)\right)
    \right]^{-1}
    \ ,
    \label{Fermi distribution of active neutrinos}
\end{aligned}
\end{equation}
where $\xi_{\nu_a}(T)\equiv\mu_{\nu_a}/T$ is the dimensionless chemical potential of the active neutrinos and is related to the lepton asymmetry as 
    \begin{equation}
        \xi_{\nu_a}(T)
        = 
        \frac{4\pi^2}{15}g_{*,s}(T) L_{\nu_a}(T)
        \ .
    \end{equation}
As we will discuss in Sec.~\ref{subsec: review of AD leptogenesis}, we consider the situation where the initial lepton asymmetry is given by $L_{\nu_e}^{\rm{init}}=L_{\nu_\mu}^{\rm{init}}=-L_{\nu_\tau}^{\rm{init}}$ or $L_{\nu_e}^{\rm{init}}=-L_{\nu_\mu}^{\rm{init}}=L_{\nu_\tau}^{\rm{init}}$. 
When the cosmic temperature is below $\mathcal{O}(10)\,{\rm{MeV}}$, the lepton asymmetry in each flavor is forced to be equal through neutrino oscillations among active neutrinos~\cite{Dolgov:2002ab,Wong:2002fa}. 
As we will see later, the resonant production of sterile neutrinos occurs at $T > \mathcal{O}(10)\,{\rm{MeV}}$ in our setup.
Therefore, in the following, we consider that only $L_{\nu_e}$ evolves in time due to sterile neutrino production while $L_{\nu_\mu}$ and $L_{\nu_\tau}$ remain constant.

The time evolution of $L_{\nu_e}$ is given by
\begin{align}
    \frac{{\rm{d}}}{{\rm{d}}t} L_{\nu_e}(T)
    =
    -\frac{45}{4\pi^4g_{*,s}(T_i)}\int{\rm{d}\epsilon}\;\epsilon^2 
    \Gamma_{\nu_e}(p,T)
    \left[
        \theta_M^2(\epsilon,T)
        f_{\nu_e}(\epsilon,T)
        -
       \bar{\theta}_M^2(\epsilon,T)
        f_{\bar{\nu}_e}(\epsilon,T)
    \right]
    \ .
    \label{eq : time evolution of lepton asymmetry}
\end{align}
We solve Eqs.~\eqref{master equation of nus distribution} and~\eqref{eq : time evolution of lepton asymmetry} simultaneously. 
As for the time dependence of the thermal width $\Gamma_{\nu_e}$, 
we use a fitting function for $y_e(T)$ obtained from the result of Refs.~\cite{Asaka:2006nq,Asaka:2006nq_web}. 
The details of our fitting function are presented in Appendix~\ref{Appendix : formula}. In the numerical calculations, we discretize $\epsilon$. 
Since the resonance occurs in a narrow range of $\epsilon$, a large number of the $\epsilon$ bins is required to assure the accuracy of the numerical calculations.
We describe the details of the discretization in Appendix~\Ref{Appendix : bin dependence}.

Since the resonant production of sterile neutrinos occurs in the radiation-dominated era, the relation between the cosmic time and the cosmic temperature is given by
\begin{equation}
    \frac{{\rm{d}}t}{{\rm{d}}T}
    =
    -\sqrt{\frac{90}{\pi^2 g_{*}(T)}}
    \frac{M_{\rm{Pl}}}{T^3}
    \left(
        1+ \frac{1}{3}\frac{{\rm{d}}\ln g_{*,s}(T)}{{\rm{d}}\ln T}
    \right)
    \ ,
\end{equation}
where $M_\mathrm{Pl} \simeq 2.4 \times 10^{18}$\,GeV is the reduced Planck mass, and $g_*$ is the number of relativistic degrees of freedom for energy.
Hereafter, we assume that $g_* = g_{*,s}$ for $T\gtrsim 1$~MeV.
Because the resonant production occurs around the QCD phase transition
in our setup, 
the time dependence of $g_{*}$ can 
affect the sterile neutrino abundance. 
We used the fitting formula of $g_*$ presented in Ref.~\cite{Wantz:2009it}.

The final density parameter of the sterile neutrino, $\Omega_{\nu_s}$, is given by
\begin{equation}
    \Omega_{\nu_s}
    =
    \frac{m_s T_0^3}{2\pi^2\rho_{{\rm{cr}},0}}\frac{g_{*,s}(T_0)}{g_{*,s}(T_i)}
    \int{\rm{d}}\epsilon\,
    \epsilon^2f_{\nu_s}(\epsilon,T=T_{*})
    \ ,
\end{equation}
where $T_0=2.73$\,K is the current cosmic temperature, $\rho_{\rm{cr},0}$ is the current critical energy density of the universe, and $T_{*}=1$\,MeV is the final temperature of the numerical calculation. 


\subsection{Analytical study}
\label{subsec : analytical study}

Before showing the results of the numerical calculation, we analytically discuss the properties of the equation~\eqref{master equation of nus distribution} that determines the evolution of the sterile neutrino density. 
In the SF mechanism, sterile neutrinos are mainly produced around the resonance point determined by 
\begin{align}
    &
    1-\frac{2\epsilon_{\rm{phys}}(T) T}{m_s^2}
    V_e(\epsilon,T) 
    =
    0
    \nonumber \\
    \Leftrightarrow ~
    &
    1-\frac{8\sqrt{2}\pi^2 g_{*}(T) G_{\rm{F}}}
    {45}
    \frac{\epsilon_{\rm{phys}}(T) L_{\nu_e} T^4}
    {m_s^2}
    +
    2 B_e
    \frac{\epsilon_{\rm{phys}}^2(T) T^6 }{m_s^2}
    =
    0
    \ .
    \label{resonance condition}
\end{align}
This equation has at most two positive solutions for $T$, which we denote by $T=T_{\rm{low}}$ and $T_{\rm{high}}$ ($T_{\rm{low}}<T_{\rm{high}}$).

Since $T_{\rm{low}}$ is approximately given by equating the first and second terms in Eq.~\eqref{resonance condition}, we find
\begin{equation}
    T_{\rm{low}}
    \propto
    \frac{m_s^{1/2}}{L_{\nu_e}^{1/4} \epsilon^{1/4}}
    \ .
    \label{input parameter dependence of resonance condition}
\end{equation}
Thus, the resonance at the lower temperature occurs from low to high $\epsilon$ modes. 
On the other hand, $T_{\rm{high}}$ is given by equating the second and third terms in Eq.~\eqref{resonance condition}, and its parameter dependence is given by
\begin{equation}
    T_{\rm{high}}
    \propto
    \frac{L_{\nu_e}^{1/2}}{\epsilon^{1/2}}
    \ .
    \label{input parameter dependence of resonance condition 2}
\end{equation}
In the non-resonant production case (the DW case), the second term in Eq.~\eqref{resonance condition} is absent, and the peak temperature of sterile neutrino production is approximately given by equating the first and third term in Eq.~\eqref{resonance condition}. Note that in general $T_{\rm{low}}$ is smaller than the peak temperature in the DW mechanism for the same value of $m_s$ and $\epsilon$. Thus, if the resonance effect plays a dominant role, the production of sterile neutrinos typically takes place at a lower cosmic temperature than in the DW case.

Neglecting the time evolution of the lepton asymmetry during the resonance of a single mode and the production from anti-electron neutrinos, the integration can be performed analytically, which gives 
\begin{align}
    f_{\nu_s} 
    \simeq
    0.96
    \frac{g_{*}(T_i)^{2/3}}
    {g_{*}(T_{\rm{low}})^{13/6} }
    \frac{M_{\rm{Pl}} m_s^4\theta^2}{G_{\rm{F}} \epsilon^2 T_{\rm{low}}^7}
    \frac{f_{\nu_e}(\epsilon,T_{\rm{low}})}{L_{\nu_e}(T_{\rm{low}})}
    \ ,
\label{analytic result of contribution from low resonance point}
\end{align}
for the resonance at $T=T_{\rm{low}}$.
For the resonance at $T=T_{\rm{high}}$, we obtain 
\begin{align}
    f_{\nu_s}
    \simeq
    1.9
    \frac{g_{*}(T_i)^{2/3}}
    {g_{*}(T_{\rm{high}})^{13/6}} 
    \frac{M_{\rm{Pl}} m_s^4 \theta^2}
    {G_{\rm{F}} \epsilon^2 T_{\rm{high}}^7}
    \frac{f_{\nu_e}(\epsilon,T_{\rm{high}})}{L_{\nu_e}(T_{\rm{high}})}
    \ .
    \label{analytic result of contribution from high resonance point}
\end{align}
See Appendix~\Ref{Appendix : approximate integration of master equation} for the derivations of these formulae. 

Since Eq.~\eqref{analytic result of contribution from low resonance point} is larger than Eq.~\eqref{analytic result of contribution from high resonance point} by a factor of $\mathcal{O}((T_{\rm{high}}/T_{\rm{low}})^7)$,
we can approximate $f_{\nu_s}$ by Eq.~\eqref{analytic result of contribution from low resonance point}
if $T_\mathrm{high}/T_\mathrm{low}$ is sufficiently large.
Then, from Eqs.~\eqref{input parameter dependence of resonance condition} and \eqref{analytic result of contribution from low resonance point}, we find
\begin{equation}
    f_{\nu_s}
    \propto
    \frac{m_s^{1/2} L_{\nu_e}^{3/4}\theta^2}{\epsilon^{1/4}}
    f_{\nu_e}
    \ .
    \label{approximate behavior of final spectrum}
\end{equation}
This indicates that the final abundance of the sterile neutrino monotonically increases with $L_{\nu_e}^{\rm{init}}$ as long as the resonance effect plays a dominant role. 
If the time evolution of $L_{\nu_e}$ is negligible, the final distribution is proportional to $\epsilon^{-1/4} f_{\nu_e}$ at $T\sim T_{\rm{low}}$, which makes the average momentum smaller than the thermal distribution. 
On the other hand, if the time evolution of $L_{\nu_e}$ is significant, the value of $L_{\nu_e}$ at the resonant production monotonically decreases for $\epsilon$ since the resonance occurs from low $\epsilon$ to high $\epsilon$.
Consequently, the final average momentum becomes further smaller than the constant $L_{\nu_e}$ case.

\subsection{Numerical results}
\label{subsec : SF numerical results}

\begin{figure}[t]
\centering
        \includegraphics[width=0.75\linewidth]{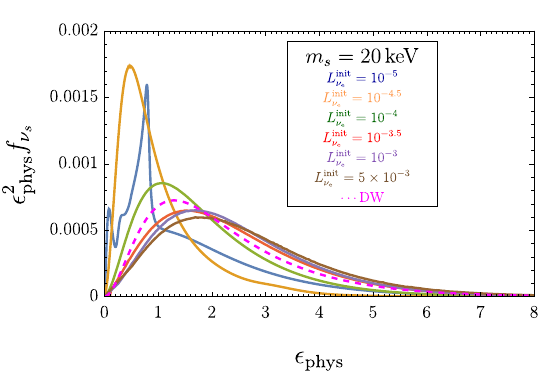}
        \\
        \includegraphics[width=0.75\linewidth]{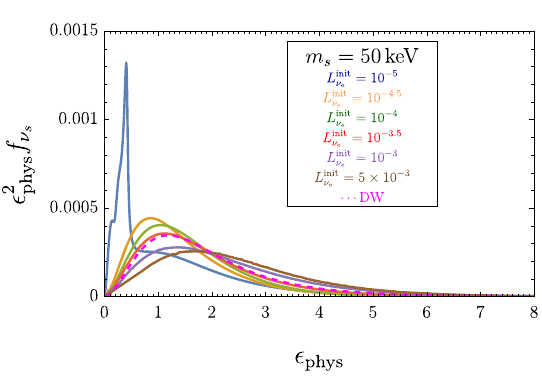}
    \caption{%
    Final spectrum of the sterile neutrino with $m_s=20$\,keV (the upper panel) and $m_s=50$\,keV (the lower panel). 
    $\epsilon_{\rm{phys}}$ is evaluated at $T=1$~MeV, and the mixing angles are fixed to explain all dark matter for each parameter set.
    The values of the initial lepton asymmetry are $L_{\nu_e}^{\rm{init}}=10^{-5},10^{-4.5},10^{-4},10^{-3.5},10^{-3}, 5\times10^{-3},0$, respectively. 
    The result for the DW mechanism ($L_{\nu_e}^\mathrm{init} = 0$) is shown in the dashed line.
    $L_{\nu_e}^{\rm{init}}=5\times 10^{-3}$ corresponds to the center value suggested by Ref.~\cite{Matsumoto:2022tlr}. 
    }
    \label{fig : final spectrum}
\end{figure}

First, we show the final spectrum of sterile neutrinos for $m_s=20$\,keV and $50$\,keV in Fig.~\Ref{fig : final spectrum}, where $\epsilon_{\rm{phys}}$ is evaluated as $T = T_*=1\,{\rm{MeV}}$.
We see that the peak momentum of the final spectrum increases as $L_{\nu_e}^{\rm{init}}$ as expected from the analytical discussion in Sec.~\Ref{subsec : analytical study}.

\begin{figure}[t]
\centering
        \includegraphics[width=0.8\linewidth]{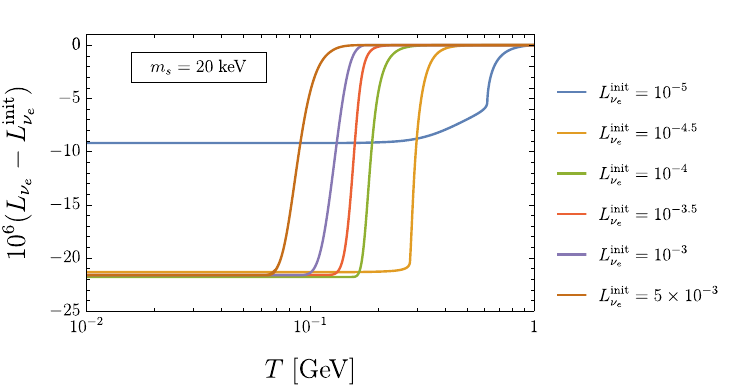}
        \\
        \includegraphics[width=0.8\linewidth]{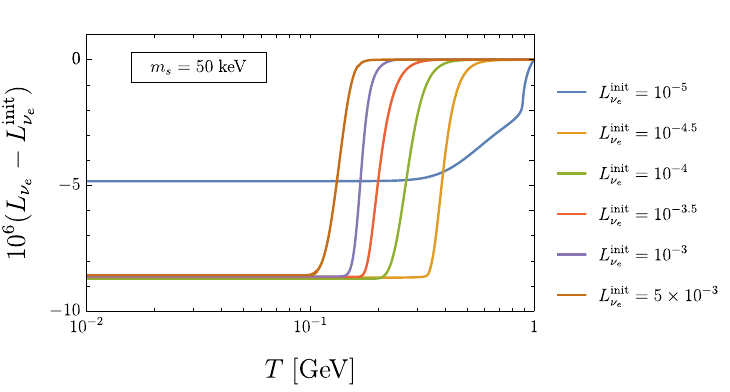}
    \caption{%
   Time evolution of lepton asymmetry with $m_s=20$\,keV (the upper panel) and  $m_s=50$\,keV (the lower panel).
   The values of the initial lepton asymmetry are $L_{\nu_e}^{\rm{init}}=10^{-5},10^{-4.5},10^{-4},10^{-3.5},10^{-3},5\times10^{-3}$, respectively. 
   Mixing angles are fixed to explain all dark matter.
   }
    \label{fig : time evolution of lepton asymmetry}
\end{figure}

We also show the time evolution of the lepton asymmetry as a function of the cosmic temperature $T$ in Fig.~\ref{fig : time evolution of lepton asymmetry}. 
We see that the lepton asymmetry is consumed around $T = \mathcal{O}(0.1\,\text{--}\,1)$\,GeV.
Furthermore, the resonance temperature decreases as $L_{\nu_e}^{\rm{init}}$ increases and $m_s$ decreases as expected from Eq.~\eqref{input parameter dependence of resonance condition}.
Note that the quantity $(n_{\nu_s}-n_{\bar{\nu_s}})/s_{\rm{tot}}+L_{\nu_e}$ is conserved.
Thus, if the resonant production is dominant, the amount of the consumed lepton asymmetry, $\Delta L_{\nu_e}\equiv L_{\nu_e}^{\rm{init}}-L_{\nu_e}(T_*)$, satisfies 
\begin{equation}
    m_{s}\Delta L_{\nu_e}
    \simeq 
    \frac{\rho_{\rm{DM},0}^{\rm{obs}}}{s_{\rm{tot},0}}
    \ ,
    \label{eq : approximation of final value of lepton asymmetry}
\end{equation}
when the sterile neutrino accounts for all dark matter.
Thus, the final values of $L_{\nu_e}-L_{\nu_e}^{\rm{init}}$ for $L_{\nu_e}^{\rm{init}} \geq 10^{-4.5}$ becomes almost the same in Fig.~\Ref{fig : time evolution of lepton asymmetry}.
On the other hand, in the case with $L_{\nu_e}^{\rm{init}}=10^{-5}$, the non-resonant production also contributes to the sterile neutrino abundance.
Therefore, the final value of $L_{\nu_e}-L_{\nu_e}^{\rm{init}}$ for $L_{\nu_e}^{\rm{init}}=10^{-5}$ is different from the other cases.

\begin{figure}[t]
\centering
\includegraphics[width=0.8\linewidth]{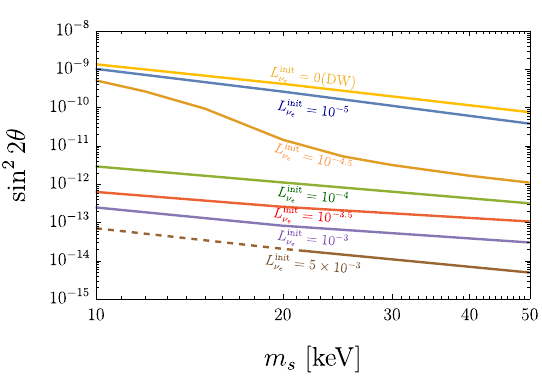}
    \caption{%
    $m_s$ and $\sin^2 2 \theta$ to explain all dark matter by the DW and SF mechanisms. 
    The dashed line denotes the parameter region where the stationary point approximation becomes unreliable
    (see Sec.~\Ref{subsec : Validity of stationary point approximation}).
    }
    \label{fig : Contour Plot}
\end{figure}

Finally, we show the contours in the $m_s$\,--\,$\sin^22\theta$ plain that explain all dark matter for a given $L_{\nu_e}^{\rm{init}}$ in Fig.~\Ref{fig : Contour Plot}. When $L_{\nu_e}^{\rm{init}}$ satisfies the condition
\begin{equation}
   m_{s}L_{\nu_e}^{\rm{init}}
   \lesssim
   \frac{\rho_{\rm{DM},0}^{\rm{obs}}}{s_{\rm{tot},0}}
   \ ,
\end{equation}
$L_{\rm{\nu_e}}$ is completely consumed and only sterile neutrinos with low momenta are resonantly produced.
After the resonant production stops, sterile neutrinos are further produced non-resonantly. 
In this case, the same order of $\theta$ as in the DW case is required to explain all dark matter since the sterile neutrinos with $\sim \rho_{\rm{DM}}^{\rm{obs}}$ must be non-resonantly produced. 
On the other hand, when $L_{\nu_e}^{\rm{init}}$ satisfies
\begin{equation}
   m_{s}L_{\nu_e}^{\rm{init}}
   \gtrsim
   \frac{\rho_{\rm{DM},0}^{\rm{obs}}}{s_{\rm{tot},0}}
   \ ,
   \label{condition for dominant resonance}
\end{equation}
almost all modes are resonantly produced and the required value of $\theta$ becomes significantly smaller than in the DW case.
Note that this behavior is different from the contours shown in Refs.~\cite{Boyarsky:2009ix,Baur:2017stq}, in which the contours for non-zero lepton asymmetry begin to deviate from that for the DW case even when $m_s$ is too small to satisfy Eq.~\eqref{condition for dominant resonance}. Although we could not specify the reason for this difference, it might come from the differences in the numerical settings such as the formalization of the mixing angle in the plasma and the assumption on the lepton asymmetry. On the other hand, we obtained the time evolution of the lepton asymmetry for $m_s=7.1$\,keV, $L_{\nu_e}^{\rm{init}} \simeq 8\times10^{-5}$, and $\sin^22\theta=7\times10^{-11}$ consistent with the numerical result in Ref.~\cite{Ghiglieri:2015jua}.

At the end of this section, we mention other proposed mechanisms for sterile neutrino dark matter production~\cite{Petraki:2007gq,Kusenko:2010ik}. 
In the model discussed in Ref.~\cite{Petraki:2007gq}, the sterile neutrinos are produced via the decay of gauge singlet Higgs bosons at $T \sim 100$\,GeV.
In the model discussed in Ref.~\cite{Kusenko:2010ik}, sterile neutrinos have a gauged $U(1)_{B-L}$ charge, and they are mainly produced from standard model particles in thermal plasma via the $U(1)_{B-L}$ gauge boson exchange.
In these models, sterile neutrinos are produced at much higher cosmic temperatures than in the resonant production model, and the averaged momentum of dark matter is more redshifted due to the time evolution of $g_*$.
Thus, these scenarios can account for sterile neutrino dark matter with masses of a few keV evading the constraint from the observations of the Lyman-$\alpha$ forests.

\subsection{Validity of the stationary point approximation}
\label{subsec : Validity of stationary point approximation}

So far, we have evaluated the production of the sterile neutrino using the stationary point approximation.
Here, we discuss the condition to apply the stationary point approximation.
When the resonance effect is neglected, Eq.~\eqref{eq : second line in density operator equation} gives a solution with $\rho_{as}$ oscillating around zero.
Since the oscillation period is typically much shorter than the cosmic expansion time scale~\cite{Dolgov:2002wy}, the right-hand side in Eq.~\eqref{eq : second line in density operator equation} is effectively considered to be zero.
Thus, we can safely use the stationary point approximation.

On the other hand, during the resonance, $\left|\frac{m_s^2}{2p}\cos2\theta-V_a\right|\ll \Gamma_{\nu_a}$ is satisfied, and Eq.~\eqref{eq : second line in density operator equation} is approximately given by
\begin{equation}
    i\frac{{\rm{d}}}{\rm{d}t} 
     \rho_{as}(t,\epsilon)
    \simeq 
    -\frac{m_s^2}{4p}\sin2\theta\rho_{aa}(t,\epsilon)
    -
    i\frac{\Gamma_{\nu_a}}{2}\rho_{as}(t,\epsilon)
    \ .
    \label{eq : approximation of density operator equation during resonance}
\end{equation}
Using $\rho_{aa}\simeq f_{\rm{eq}}$, we can approximate that the first term in Eq.~\eqref{eq : approximation of density operator equation during resonance} is time-independent and the solution is given by $\rho_{as}\propto 1-\exp(-\Gamma_{\nu_a} t/2)$. 
Then, we regard that the left-hand side of Eq.~\eqref{eq : approximation of density operator equation during resonance} converges to zero in time scale $\sim \Gamma_{\nu_a}^{-1}$.
Thus, we consider that if the resonance width $\delta t_{\rm{res}}$ satisfies $\delta t_{\rm{res}}\gg \Gamma_{\nu_a}^{-1}$, the stationary point approximation can be safely used. 

\begin{figure}[t]
\centering
\includegraphics[width=0.8\linewidth]{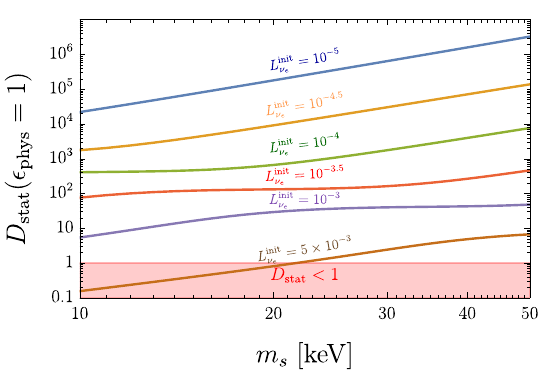}
    \caption{%
    $D_{\rm{stat}}(\epsilon_{\rm{phys}}=1)$ defined in Eq.~\eqref{eq : during resonance condition4}.
    For $D_{\rm{stat}}(\epsilon_{\rm{phys}}=1) < 1$, the stationary point approximation becomes unreliable.
    }
    \label{fig : stationary point approximation}
\end{figure}

Now let us evaluate the resonance width $\delta t_{\rm{res}}$.
The condition for the resonance is given by 
\begin{equation}
    \left| 1- \frac{2p}{m_s^2}V_a \right|
    \lesssim 
    \frac{p}{m_s^2}\Gamma_{\nu_a}
    \ .
    \label{eq : during resonance condition}
\end{equation}
Then, the temperature interval satisfying Eq.~\eqref{eq : during resonance condition}, $\delta T_{\rm{res}}$, is given by
\begin{equation}
    \frac{\delta T_{\rm{res}}}{T_{\rm{res}}}
    \sim
    \frac{\Gamma_{\nu_a}(T_{\rm{res}})}{3V_a(\epsilon_{\rm{phys}},T_{\rm{res}})}
    \ .
    \label{eq : during resonance condition2}
\end{equation}
Therefore, the duration time of resonance $\delta t_{\rm{res}}$ is given by
\begin{equation}
    \delta t_{\rm{res}}
    \sim
    \left.
        \frac{{\rm{d}}t}{{\rm{d}}T}
    \right|_{T_{\rm{res}}}
    \times 
    \frac{\Gamma_{\nu_a}(T_{\rm{res}})T_{\rm{res}}}{3V_a(\epsilon_{\rm{phys}},T_{\rm{res}})}
    \ .
    \label{eq : during resonance condition3}
\end{equation}
Here, we regard that the stationary point approximation works if $\delta t_{\rm{res}}> \Gamma_{\nu_a}^{-1}$.
To quantify the validity of the stationary point approximation, we define
\begin{equation}
    D_{\rm{stat}}
    \equiv
    \left.
        \frac{{\rm{d}}t}{{\rm{d}}T}
    \right|_{T_{\rm{low}}}
    \times
    \frac{\Gamma_{\nu_a}^2(T_{\rm{low}})T_{\rm{low}}}
    {3V_a(\epsilon_{\rm{phys}}, T_{\rm{low}})}
    \ ,
    \label{eq : during resonance condition4}
\end{equation}
so that the condition to use Eq.~\eqref{master equation of nus distribution} is given by $D_{\rm{stat}}>1$.
Here we evaluate the condition at the lower resonance temperature $T_\mathrm{low}$ when the sterile neutrino production is dominant.
We show $D_{\rm{stat}}$ as a function of $m_s$ for $\epsilon_{\rm{phys}}=1$ and several values of $L_{\nu_e}$ in Fig.~\ref{fig : stationary point approximation}. 
We find that the stationary point approximation is valid for $m_s > 10$\,keV with $L_{\nu_e}\lesssim 10^{-3}$. 
On the other hand, it is valid for $m_s\gtrsim 20$\,keV in the case with $L_{\nu_e}^{\rm{init}}=5\times10^{-3}$.

\section{Large lepton asymmetry from Affleck-Dine mechanism}
\label{sec: AD leptogenesis}

In this section, we discuss the scenario of Affleck-Dine (AD) leptogenesis~\cite{Affleck:1984fy,Dine:1995kz,Murayama:1993em} to realize the resonant production of sterile neutrinos.

\subsection{Affleck-Dine leptogenesis}
\label{subsec: review of AD leptogenesis}

AD leptogenesis utilizes flat directions in the scalar potential of the minimal supersymmetric standard model (MSSM). 
Here, we consider one of the leptonic flat directions and call it the AD field.
The potential of the AD field is lifted by supersymmetry (SUSY) breaking effects and the existence of a cutoff at the Planck scale, $M_\text{Pl}$.

During inflation, the AD field has a potential given by
\begin{align}
    V(\phi) 
    & =
    m_\phi^2|\phi|^2 - cH^2|\phi|^2
    + V_\text{NR} + V_\text{A}
    \nonumber\\
    & =
    m_\phi^2|\phi|^2 - cH^2|\phi|^2
    + |\lambda|^2 \frac{|\phi|^{2(n-1)}}{M_{\rm{Pl}}^{2(n-3)}} 
    + \left(\lambda a_M \frac{m_{3/2}\phi^n}{n M_{\rm{Pl}}^{n-3}}
    + \mathrm{h.c.}\right)
    \ ,
    \label{AD field potential}
\end{align}
where $m_\phi$ is the soft SUSY breaking mass of the AD field, $m_{3/2}$ is the gravitino mass, and $c$, $a_M$, and $\lambda$ are dimensionless constants. 
In gravity-mediated SUSY breaking models, $m_{\phi}\sim m_{3/2}$ while $m_\phi$ is generally much larger than $m_{3/2}$ for gauge-mediated SUSY breaking models.\footnote{
More precisely, in gauge-mediation models $m_\phi$ is much larger than $m_{3/2}$ for $|\phi|$ less than the messenger scale.
For large field values, the soft SUSY breaking mass term is more complicated and the effective mass decreases with $|\phi|$ until it reaches $m_{3/2}$ as shown in Sec.~\ref{subsec: delayed type}.}
The third term $V_\text{NR}$ is called the non-renormalizable term, and the fourth term $V_\text{A}$ is the A-term, which violates a global $U(1)$ symmetry associated with lepton number.
$n\, (\ge 4)$ is an integer depending on the choice of the flat direction and non-renormalizable superpotential.
Here, we assume $c=\mathcal{O}(1)>0$.

After inflation, $H$ decreases and becomes smaller than the SUSY breaking mass at some point.
While $H\gtrsim m_{\phi}$, the AD field has a non-zero field value given by
\begin{equation}
    |\phi|
    \simeq
    \left(
        \frac{HM_{\rm{Pl}}^{n-3}}{|\lambda|}
    \right)^{\frac{1}{n-2}}
    \ ,
\end{equation}
due to the balance between the Hubble-induced mass and the non-renormalizable term.
When $H$ becomes comparable to $m_{\phi}$, the origin of the field space is stabilized due to the soft mass term, and the AD field starts to oscillate around the origin. 
At this period, the field value is given by 
\begin{equation}
    |\phi|
    \simeq 
    \varphi_{\rm{osc}}
    \equiv 
    \left(
        \frac{m_{\phi} M_{\rm{Pl}}^{n-3}}{|\lambda|}
    \right)^{\frac{1}{n-2}}
    \ .
\end{equation}
At the same time, the A-term kicks the AD field in the phase direction.
Consequently, the AD field acquires a lepton number proportional to $\varphi^2 \dot{\theta}_\mathrm{AD}$, where $\varphi$ and $\theta_\mathrm{AD}$ are the radial and phase components of $\phi$, i.e., $\phi \equiv \varphi e^{i\theta_\mathrm{AD}}$. 
The resulting lepton number density
is written as
\begin{equation}
    n_L(t_\mathrm{osc}) \simeq \delta\, m_\phi \varphi_\mathrm{osc}^2 
    \ ,
    \label{resulting lepton number in AD leptogenesis}
\end{equation}
where $\delta$ is the efficiency parameter depending on $n$, the initial phase of the AD field, and the phases of $a_M$ and $\lambda$.
Notice that $\delta$ can take a positive or negative value depending on the initial phase of the AD field.

Now, we comment on the flat direction considered in our scenario. 
As a concrete flat direction, we consider $\mu^{c}L_eL_\tau$ or $\tau^{c}L_eL_\mu$ direction,
where $a^c$ is the charge conjugation of an SU(2)$_{\rm{W}}$ singlet charged lepton with flavor $a =e, \mu, \tau$, and
$L_a$ is an SU(2)$_{\rm{W}}$ doublet lepton~\cite{Gherghetta:1995dv}. 
As a result, the generated lepton asymmetry satisfies $L_{\nu_e}=-L_{\nu_\mu}=L_{\nu_\tau}$ or $L_{\nu_e}=L_{\nu_\mu}=-L_{\nu_\tau}$, respectively.
Then, the first term in the right-hand side in Eq.~\eqref{eq : finite density potential} is equal to $2\sqrt{2}G_{\rm{F}}L_{\nu_e}s_{\rm{tot}}$, which can induce the resonant production of sterile neutrinos.\footnote{On the other hand, $L=0$ for $e^{c}L_\mu L_\tau$ flat direction.}  

\subsection{Delayed-type L-ball scenario}
\label{subsec: delayed type}

From now on, we consider the gauge-mediated SUSY breaking scenario.
After the AD field starts to oscillate, the AD field potential is given by
\begin{align}
    V
    &=
    V_{\rm{gauge}} + V_{\rm{gravity}}
    \ .
    \label{gauge mediation and gravity mediation potential}
\end{align}
Here, $V_{\rm{gauge}}$ is the contribution from gauge-mediation given by~\cite{deGouvea:1997afu,Kusenko:1997si}
\begin{align}
    V_{\rm{gauge}}
    =
    M_{\rm{F}}^4 \left( \ln \frac{|\phi|^2}{M_S^2} \right)^2
    \ ,
    \label{eq:Vgauge}
\end{align}
where $M_{\rm{F}}$ is the SUSY breaking scale, and $M_S$ is the scale of the messenger sector.
Here, we assumed $|\phi|\gg M_S$.
$V_{\rm{gravity}}$ is the contribution from gravity-mediation given by~\cite{deGouvea:1997afu,Kusenko:1997si}
\begin{align}
    V_{\rm{gravity}}
    =
    m_{3/2}^2
    \left[
        1 + K\ln\left( \frac{|\phi|^2}{M_{\rm{Pl}}^2} \right)
    \right]
    |\phi|^2
    \ ,
    \label{eq:Vgravity}
\end{align}
where $K$ is a dimensionless constant, which we assume to be positive in this paper.
We also assume that $V_{\rm{gravity}}$ is dominant in the potential when the AD field starts to oscillate, i.e., $\varphi_\mathrm{osc} > \varphi_{\rm{eq}}\equiv M_{\rm{F}}^2/m_{3/2}$.
When the potential is dominated by $V_{\textrm{gravity}}$ with $K>0$, the homogeneous oscillation of the AD field is stable.
However, right after $\varphi$ becomes comparable to $\varphi_{\rm{eq}}$, $V_{\rm{gauge}}$ controls the dynamics of the AD field.
In this period, the AD field experiences spatial instabilities and forms spherically symmetric non-topological solitons called L-balls~\cite{Enqvist:1997si,Kasuya:1999wu,Kasuya:2014ofa}.
We call this type of L-ball formation scenario ``delayed-type'' L-ball scenario~\cite{Kasuya:2001hg}.

Delayed-type L-balls have the same properties as the gauge-mediated type L-balls.
Then, the initial charge, mass, radius, and energy per charge of a delayed-type L-ball are given by~\cite{Kasuya:2014ofa}
\begin{align}
    |Q_{\rm{G}}^{\mathrm{init}}|
    & \simeq
    \beta_{\rm{G}} \frac{\varphi_{\rm{eq}}^4}{M_{\rm{F}}^4}
    \simeq
    \beta_{\rm{G}} 
    \left(
        \frac{M_{\rm{F}}}{m_{3/2}}
    \right)^4
    \ , 
    \label{Q_G}
    \\[0.4em]
    M_Q & \simeq \frac{4\sqrt{2}\pi}{3}\zeta M_{\rm{F}} 
    |Q_{\rm{G}}|^{\frac{3}{4}}
    \ ,
    \label{eq : Q-ball property1}
    \\[0.4em]
    R_Q & \simeq \frac{1}{\sqrt{2}\zeta}M_{\rm{F}}^{-1}
    |Q_{\rm{G}}|^{\frac{1}{4}}
    \ ,
    \label{eq : Q-ball property2}
    \\[0.4em]
    \omega_Q & \simeq \sqrt{2}\pi \zeta M_{\rm{F}}
    |Q_{\rm{G}}|^{-\frac{1}{4}}
    \ ,
    \label{eq : Q-ball property3}
\end{align}
where $\beta_{\rm{G}}\simeq 6\times10^{-5}$~\cite{Kasuya:2012mh} in the case with $|\delta| \lesssim 10^{-1}$, which we consider in this paper,
and $\zeta$ is a numerical factor  $\simeq 2^{1/4}(c_0/\pi)^{1/2}$ where $c_0\equiv 4.8\ln (m_\phi/\sqrt{2}\omega_Q)+7.4$~\cite{Hisano:2001dr}. In the following, we adopt $\zeta=3.4$, which is obtained by substituting $m_{3/2}=1$\,GeV, $M_{\rm{F}}=10^6$\,GeV and $m_\phi=10^4$\,GeV. 
The L-balls decay into neutrinos with the decay rate $\Gamma_{Q}$ given by~\cite{Cohen:1986ct,Kawasaki:2012gk}
\begin{equation}
    \Gamma_{Q}
    \equiv 
    -\frac{1}{Q_\mathrm{G}} \frac{\mathrm{d}Q_\mathrm{G}}{\mathrm{d} t}
    \simeq 
    \frac{N_l}{|Q_\mathrm{G}|}\frac{\omega_Q^3}{12\pi^2}4\pi R_Q^2
    \simeq
    \frac{\sqrt{2}\pi^2 N_l\zeta}{3\beta^{5/4}}\frac{m_{3/2}^5}{M_{\rm{F}}^4}
    \ ,
    \label{Qball decay rate}
\end{equation}
where $N_l=3$ is the number of active neutrino species.

To realize the generation of a large lepton asymmetry from L-ball decay, we assume that L-balls dominate the energy density of the universe at L-ball decay. 
Then, the resulting lepton-to-entropy ratio from L-ball decay is evaluated as
\begin{align}
    L_{\nu_e}^{\rm{init}} &\simeq \frac{n_L(t_{\rm{osc}})}{m_{3/2}^2\varphi_{\rm{osc}}^2} 
    \frac{\rho_{\rm{Q}}(t_{\rm{decay}})}{s(t_{\rm{decay}})} 
    \nonumber\\
    &\simeq
    \delta\frac{3T_{\rm{decay}}}{4m_{3/2}}
    \ ,
    \label{Relation between etaL gravitino mass and Tdecay}
\end{align}
where $\rho_{\rm{Q}}$ is the energy density of the L-balls, and $T_\mathrm{decay}$ is the cosmic temperature just after L-ball decay. 
L-balls decay almost instantaneously when the decay rate $\Gamma_Q$ becomes equal to the Hubble parameter. 
Thus, $T_\mathrm{decay}$ is given by
\begin{align}
    T
    \simeq
    T_{\rm{decay}} 
    &\equiv 
    \left(\frac{90}{\pi^2 g_{\ast}(T_{\rm{decay}})}\right)^{1/4}\sqrt{M_{\rm{Pl}}\Gamma_{Q}}\nonumber
    \\
    &\simeq
    2.7\,{\rm{GeV}}\times \left(\frac{g_{\ast}}{80}\right)^{-1/4}
    \left(\frac{m_{3/2}}{1\,{\rm{GeV}}}\right)^{5/2}\left(\frac{M_{\rm{F}}}{10^6\,{\rm{GeV}}}\right)^{-2}
    \ .
    \label{Tdecay}
\end{align}

Finally, we discuss the condition for L-balls to dominate the universe before the L-ball decay.
To this end, we define a ratio $f_{\rm{Q}}$ of the energy density of L-balls (or the AD field) $\rho_{\rm{Q}}$ to that of the inflaton and inflaton decay products $\rho_{\rm{R}}$.
Considering that $f_{\rm{Q}}$ is almost constant during $t_{\rm{osc}}\lesssim t\lesssim t_{\rm{R}}$, where $t_{\rm{R}}$ is the cosmic time at the completion of reheating, and scales as 
$\propto T^{-1}$ after reheating,
we obtain 
\begin{align}
    f_{\rm{Q}}(T_{\rm{decay}})
    \simeq
    \frac{g_*(T_{\rm{decay}})T_{\rm{decay}}^4}{g_*(T_{\rm{relic,dec}})T_{\rm{relic,dec}}^4}
    \simeq \frac{m_{3/2}^2\varphi_{\rm{osc}}^2}{3M_{\rm{Pl}}^2H_{\rm{osc}}^2}
    \frac{T_\mathrm{R}}{T_\mathrm{relic,dec}}
    \ ,
    \label{determining a/aeq1 1}
\end{align}
where $T_{\rm{relic,dec}}$ is the temperature of relic plasma from the inflaton right before the L-ball decay, and $T_{\rm{R}}$ is the reheating temperature.
Then, we obtain
\begin{align}
    f_{\rm{Q}}(T_{\rm{decay}})
    \simeq
    \left(
    \frac{g_*(T_{\rm{relic,dec}})}
    {g_*(T_{\rm{decay}})}
    \right)^{1/3}
    \left( \frac{T_{\rm{R}}}{T_{\rm{decay}}} \right)^{4/3}
    \left( \frac{\varphi_{\rm{osc}}^2}{3M_{\rm{Pl}}^2} \right)^{4/3}
    \ .
    \label{determining a/aeq1 4}
\end{align}
Thus, the condition $f_{\rm{Q}}(T_{\rm{decay}})>1$ can be satisfied for a sufficiently large $\varphi_{\rm{osc}}(\lesssim M_{\rm{Pl}})$ and $T_{\rm{R}}>T_{\rm{decay}}$.

Now, we search for the compatible parameter space to realize the resonant production of sterile neutrinos.
When the decay temperature of L-balls is lower than the resonance temperature of sterile neutrinos, the resonant production does not occur.
Therefore, we require $T_{\rm{low}}<T_{\rm{decay}}$ for $\epsilon$ with which the sterile neutrinos are efficiently produced.
In the following, we fix $\epsilon_{\rm{phys}}=1$ as a typical value and evaluate $T_{\rm{low}}$ as a function of the sterile neutrino mass $m_s$ and the initial lepton asymmetry $L_{\nu_e}^{\rm{init}}$.
We denote the resulting value of $T_{\rm{low}}$ as $T_{\rm{low}}(m_s,L_{\nu_e}^{\rm{init}})$.
Consequently, the condition for the resonant production is written as
\begin{equation}
    T_{\rm{low}}
    \left(
        m_s,L_{\nu_e}^{\rm{init}} = \delta\frac{3T_{\rm{decay}}}{4m_{3/2}}
    \right)
    <
    T_{\rm{decay}}
    \ .
    \label{compatible condition for resonance}
\end{equation}
Using Eqs.~\eqref{Tdecay} and~\eqref{compatible condition for resonance}, we obtain the condition that $m_{3/2}$, $M_{\rm{F}}$, and $m_s$ should satisfy as
\begin{align}
    M_{\rm{F}}  \lesssim & 
    ~ 3.1\times10^6\,{\rm{GeV}}
    \left(
    \frac{g_{*}(T_{\rm{decay}})}{80}
    \right)^{-1/8}
    \left(
    \frac{g_{*}(T_{\rm{low}})}{60}
    \right)^{1/10}
    \left(
    \frac{|\delta|}{10^{-3}}
    \right)^{1/10} \nonumber \\
    & ~ \times \left(
    \frac{m_{3/2}}{1\,{\rm{GeV}}}
    \right)^{23/20}
    \left(
    \frac{m_{s}}{50\,{\rm{keV}}}
    \right)^{-1/5}.
    \label{compatible condition for resonance in terms of MF}    
\end{align}

Next, we consider the baryon asymmetry produced from the evaporation process of L-balls until the sphaleron process becomes out of equilibrium~\cite{Laine:1998rg}. The total evaporated baryon charge can be written as~\cite{Kasuya:2014ofa}%
\footnote{
Here, we assume $T_{\rm{ew}}<T_{\rm{eq}}$.
}

\begin{align}
   \frac{\Delta Q_b}{Q_{\rm G}^\mathrm{init}}
    \simeq 
    \begin{cases}
    -
    \frac{16\pi}{23\sqrt{5}}
    \frac{ \sqrt{g_*(T_{\rm{R}})} }{ g_*(T_{\rm{ew}}) }
    \left(
        32 + 96 \log \frac{T_{\rm{eq}}}{T_{\rm{ew}}}
    \right)
    \frac{M_{\rm{Pl}}T_{\rm{R}}^2}{\zeta M_s^2M_F}
    |Q_{\rm G}^\mathrm{init}|^{-3/4} 
    &
    (T_{\rm{R}}<T_{\rm{ew}} < T_\mathrm{eq})
    \vspace{1mm} \\ 
    -\frac{16\pi }{23\sqrt{5}}\frac{1}{\sqrt{g_*(T_{\rm{ew}})}}
    \left(
        47
        - 15\frac{T_\mathrm{ew}^2}{T_\mathrm{R}^2}
        +96\log\frac{T_{\rm{eq}}}{T_\mathrm{R}}
    \right)
    \frac{M_\mathrm{Pl} T_\mathrm{R}^2}
    {\zeta M_s^2M_{\rm{F}}}
    |Q_{\rm G}^\mathrm{init}|^{-3/4}
    &
    (T_\mathrm{ew} < T_\mathrm{R} < T_{\rm{eq}})
    \vspace{1mm} \\
    -\frac{16\pi }{23\sqrt{5}}\frac{1}{\sqrt{g_*(T_{\rm{ew}})}}
    \left(
        45 - 15\frac{T_\mathrm{ew}^2}{T_{\rm{eq}}^2}
        + 2 \frac{T_\mathrm{eq}}{T_\mathrm{R}}
    \right)
    \frac{M_\mathrm{Pl} T_{\rm{eq}}^2}{\zeta M_s^2M_{\rm{F}}}
    |Q_{\rm G}^\mathrm{init}|^{-3/4}
    &
    (T_\mathrm{ew} < T_{\rm{eq}} < T_\mathrm{R})
    \end{cases}
    \ ,
    \label{The total evaporation charge}
\end{align}
where $M_s$ is the mass of gauginos involved in the evaporation process.

Here, $T_{\rm{eq}}\sim (M_s^2M_{\rm{F}})^{1/3}
|Q_G^\mathrm{init}|^{-1/12}$
is the temperature at which the evaporation rate of L-balls~\cite{Laine:1998rg} and the diffusion rate of leptons evaporated from L-balls~\cite{Banerjee:2000mb} becomes equal, and $T_{\rm{ew}}$ is the temperature at electroweak phase transition. We assumed $g_*(T)=g_*(T_{\rm{ew}})$ for $T>T_{\rm{ew}}$.
Then, if $T_\mathrm{R} < T_\mathrm{ew} < T_\mathrm{eq}$, the total baryon asymmetry induced from the L-ball evaporation process can be evaluated as
\begin{align}
    \eta_b &\simeq \eta_L\left(\frac{\Delta Q_b}{Q_{\rm{G}}^{\rm init}}\right) 
    \nonumber \\
    &\simeq 
    -
    5.5\times 10^{-11}
    \left(\frac{g_*}{80}\right)^{-\frac{1}{4}}
    \left(\frac{\delta}{10^{-3}}\right)
    \left(\frac{M_s}{10^4\,{\rm{GeV}}}\right)^{-2}
    \left(\frac{m_{3/2}}{1\,{\rm{GeV}}}\right)^{\frac{9}{2}}
    \left(\frac{M_{\rm{F}}}{10^6\,{\rm{GeV}}}\right)^{-6}
    \left(\frac{T_{\rm{R}}}{10\,{\rm{GeV}}}\right)^{2},
    \label{Etab}
\end{align}
where $\eta_L$ is the lepton-to-entropy ratio and we take $\eta_L=L_{\nu_e}^{\mathrm{init}}$. 
We require that this resulting baryon asymmetry does not spoil the BBN, that is, $|\eta_b|<|\eta_b^{\rm{obs}}|\sim 10^{-10}$, which leads to the upper bound of $T_{\rm{R}}$ in terms of $M_s,m_{3/2}$, and $M_{\rm{F}}$.
On the other hand, the L-ball domination ($f_\mathrm{Q} \gg 1$) requires $T_\mathrm{R} \gtrsim T_\mathrm{decay}$.
Thus, using Eq.~\eqref{Tdecay} and the upper bound on $T_\mathrm{R}$, we obtain 

\begin{align}
    M_{\rm{F}}  \gtrsim  7.3\times10^5\,{\rm{GeV}}
    \left(
    \frac{g_{*}(T_{\rm{decay}})}{80}
    \right)^{-3/40}
    \left(
    \frac{|\delta|}{10^{-3}}
    \right)^{1/10} 
    \left(
    \frac{m_{3/2}}{1\,{\rm{GeV}}}
    \right)^{19/20}
    \left(
    \frac{M_{s}}{10^4\,{\rm{GeV}}}
    \right)^{-1/5}.
    \label{compatible condition for baryon number}    
\end{align}

Notice that if $\eta_L <0$ and $\eta_b \sim 10^{-10}$, the baryon asymmetry produced through the L-ball evaporation can explain the present baryon asymmetry of the universe although negative lepton asymmetry may not be favored by the recent observation of the primordial helium abundance~\cite{Matsumoto:2022tlr}.

Finally, we consider the gravitino problem. The gravitino density parameter $\Omega_{3/2}$ is approximately given by~\cite{Kawasaki:2017bqm}
\begin{align}
    \Omega_{3/2}h^2
    &\simeq 
    0.71\left(\frac{m_{3/2}}{0.5\,{\rm{GeV}}}\right)^{-1}\left(\frac{M_{\tilde{g}}}{10^4\,{\rm{GeV}}}\right)^2\left(\frac{T_{\rm{R}}}{10^5\,{\rm{GeV}}}\right)
    s_Q^{-1}
    \ ,
\end{align} 
where $M_{\tilde{g}}$ is the gluino mass. 
Here, we added a factor $s_Q$, which represents the dilution of gravitinos due to entropy production via L-balls decay and is defined by
\begin{align}
    s_Q
    \equiv
    \frac{g_*(T_{\rm{decay}})T_{\rm{decay}}^3}{g_*(T_{\rm{relic,dec}})T_{\rm{relic,dec}}^3}
    \ .
    \label{eq : definition of sQ}
\end{align} 
From Eq.~\eqref{determining a/aeq1 1}, we obtain
\begin{align}
    s_Q
    \sim f_Q(T_{\rm{decay}})^{3/4}
    \ .
\end{align} 
The condition that $\Omega_{3/2}h^2\ll 0.12$ can be satisfied if we assume sufficiently large $\varphi_{\rm{osc}}\;(\lesssim M_{\rm{Pl}})$ in the parameter region of interest.
Although gravitinos are also generated from L-ball decay, its contribution is negligibly small compared with the thermal contribution~\cite{Kasuya:2012mh}.

\begin{figure}[t]
\centering
    \begin{tabular}{c}
        \includegraphics[width=0.48\linewidth]{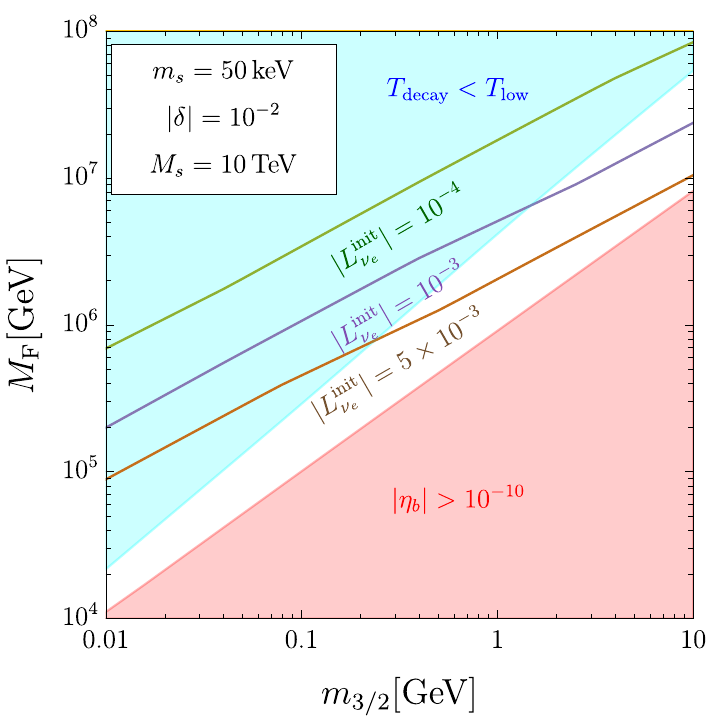}
        \includegraphics[width=0.48\linewidth]{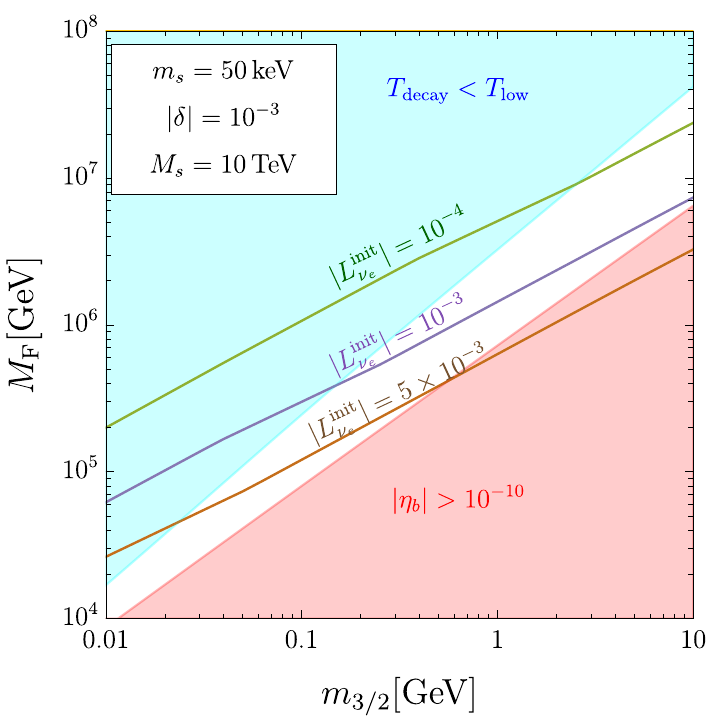}
        \\
        \includegraphics[width=0.48\linewidth]{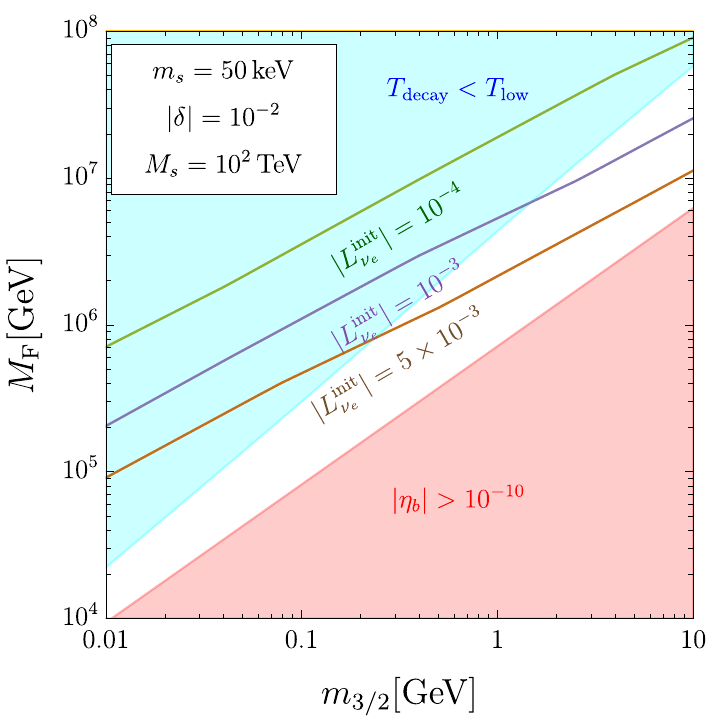}
        \includegraphics[width=0.48\linewidth]{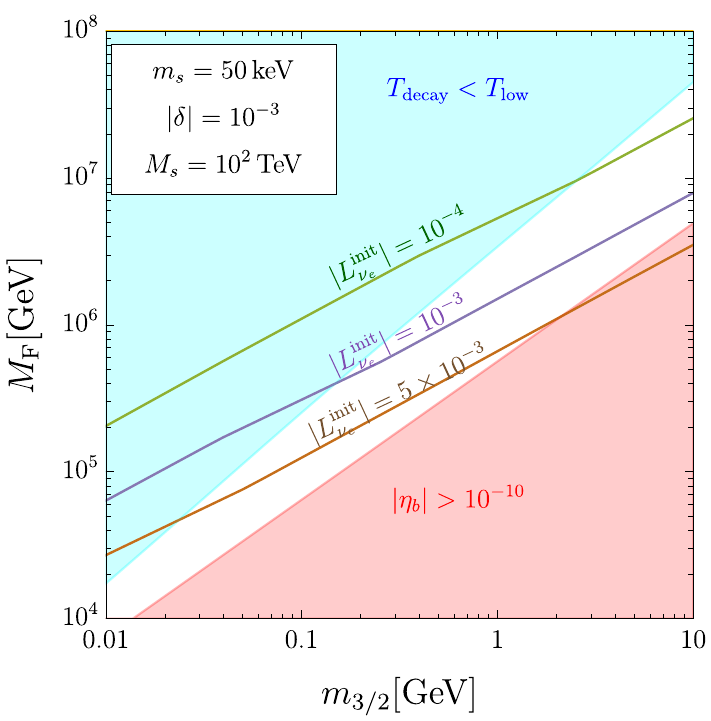}
    \end{tabular}
    \caption{%
    Constraints on $m_{3/2}$ and $M_\mathrm{F}$ in the delayed-type L-ball scenario.
    We set $M_s=10\,{\rm{TeV}}$ in the upper panels and $M_s=10^2\,{\rm{TeV}}$ in the lower panels. 
    We set $\delta=10^{-2}$ and $\delta=10^{-3}$ in the left and right panels, respectively.
    The green, violet, and brown lines correspond to $|L_{\nu_e}^\mathrm{init}|=10^{-4}$, $10^{-3}$, and $5\times10^{-3}$, respectively. 
    The red-shaded region is excluded because baryon asymmetry is overproduced,
    $|\eta_b|>10^{-10}$,
    due to the evaporation process of L-balls when we require $T_{\rm{R}}>T_{\rm{decay}}$ (see Eq.~\eqref{compatible condition for baryon number}).
    The cyan region is excluded because $T_{\rm{decay}}$ is lower than $T_{\rm{low}}$ (see Eq.~\eqref{compatible condition for resonance in terms of MF}).}
    \label{fig : Constraints on m32 and MF with mphi 10TeV}
\end{figure}

We show the constraints on $m_{3/2}$ and $M_{\rm{F}}$ for fixed values of $m_s,|\delta|$, and $M_s$ in Fig.~\Ref{fig : Constraints on m32 and MF with mphi 10TeV}. 
The red region with large $m_{3/2}$ and small $M_{\rm{F}}$ is excluded by the constraint from $|\eta_b|$ generated via L-ball evaporation process and sphaleron process.
On the other hand, the cyan region with small $m_{3/2}$ and large $M_{\rm{F}}$ is excluded by the requirement that the decay temperature of L-balls should be higher than the resonance temperature of sterile neutrino production. 
The parameters realizing $|L_{\nu_e}^{\rm{init}}|\lesssim 10^{-4}$ is severely constrained by the latter condition because the resonance temperature becomes higher with a lower value of $|L_{\nu_e}^{\rm{init}}|$ (see Eq.~\eqref{resonance condition}).
On the other hand, the parameters realizing $|L_{\nu_e}^{\rm{init}}|\gtrsim 10^{-4}$ are allowed in a wide range.

\section{Observational implications}
\label{sec: observational implications}

In this section, we discuss the observational implications of our scenario.

\subsection{Free-streaming length of sterile neutrinos}
\label{subsec: free streaming length}

Here, we discuss the free streaming of the produced sterile neutrinos. 
From the observations of the Lyman-$\alpha$ forests in the quasar spectra, the free streaming of dark matter can be constrained. 
This constraint is highly dependent on the final spectrum of sterile neutrinos. 
Here, we obtain a constraint on the mass of sterile neutrinos with the vacuum mixing angle $\theta$ fixed to explain all dark matter. 
This constraint is discussed in detail by Ref.~\cite{Baur:2017stq}. 
However, as we mentioned in Sec.~\Ref{subsec : SF numerical results}, our relation among $\theta$, $m_s$, and $L_{\nu_e}^{\rm{init}}$ to explain all dark matter is different from one used in Ref.~\cite{Baur:2017stq}. 
Thus, we revisit the constraint here.
In deriving the constraint, we first use the Lyman-$\alpha$ constraint on the mass of early decoupled thermal relics~\cite{Baur:2015jsy} and translate it into the upper bound of the free-streaming length $\lambda_{\rm{FS}}^{\rm{upper}}$ of dark matter.
We then estimate the free-streaming lengths from the final spectra obtained in Sec.~\Ref{sec: calculation of Shi Fuller mechanism} and compare them to $\lambda_{\rm{FS}}^{\rm{upper}}$.
Finally, we obtain the lower bound on $m_s$ for each value of $L_{\nu_e}^{\rm{init}}$
in a similar way to Ref.~\cite{Gelmini:2019wfp}.

The typical free-streaming length of particles decoupled from the thermal plasma is given by~\cite{Baur:2017stq}
\begin{align}
   \lambda_{\rm{FS}}
   \simeq 
   a(t)\int_{a_\mathrm{in}}^{a(t)}{\rm{d}}a\frac{v(\langle\epsilon\rangle,t)}{a(t)^2H(t)}
   \ ,
   \label{eq : free streaming length}
\end{align} 
where $\langle \epsilon \rangle$ is $\epsilon$ averaged over the momentum distribution
at $T = T_\mathrm{in}$, which is the lower one of the decoupling temperature and the particle production temperature,
and $a_\mathrm{in}$ is the scale factor at $T = T_\mathrm{in}$.

We evaluate the free-streaming length of early decoupled thermal relics as a function of the relic mass $m_x$. 
Here, an early decoupled thermal relic follows the Fermi-Dirac distribution given by
\begin{align}
    f_X(p)  
    &=
    \frac{1}{1+\exp(p/T_x)}
    \ ,
    \label{eq : relic distribution}
\end{align}
where $T_x$ is the temperature of the thermal relic, and its energy density is equal to $\rho_{\rm{DM}}^{\rm{obs}}$. 
Therefore, the relation between $m_x$ and $T_x$ after neutrino decoupling is given by~\cite{Baur:2015jsy,Gelmini:2019wfp,Viel:2005qj}
\begin{align}
   \frac{m_{\rm{eff}}}{m_x}=\left(\frac{T_x}{T_\nu}\right)^3,
   \label{eq : relation between relic temperature and relic mass}
\end{align}
where $m_{\rm{eff}}=94\,{\rm{eV}} \times \Omega_{\rm{DM}}h^2$ is the total mass of the active neutrinos for their energy density to be the same as dark matter, and $T_\nu$ is the temperature of active neutrinos.
Using the relation between $m_x$ and $\lambda_\mathrm{FS}$, we translate the lower bound, $m_x>4.09$\,keV ($95\%$ CL)~\cite{Baur:2015jsy}, into the upper bound of $\lambda_{\rm{FS}}$, $\lambda_{\rm{FS}} < \lambda_{\rm{FS}}^{\rm{upper}}=0.099$\,Mpc.

\begin{figure}[t]
\centering
\includegraphics[width=0.7\linewidth]{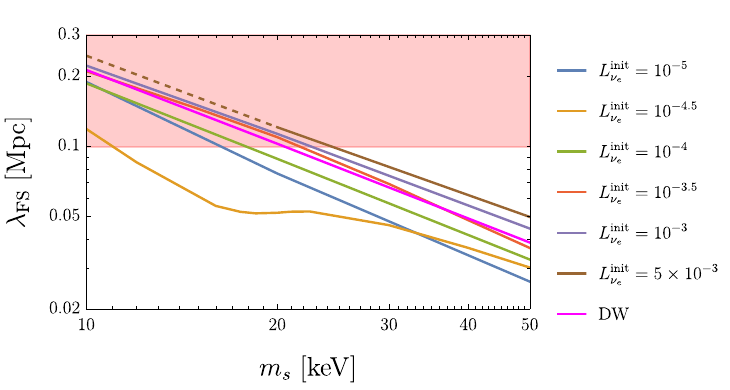}
    \caption{%
    Free-streaming length of sterile neutrinos $\lambda_\mathrm{FS}$ as a function of $m_s$.
    The red-shaded region is constrained by the Lyman-$\alpha$ forest observations.
    }
    \label{fig : Free streaming length}
\end{figure}

In the case of the sterile neutrino, we take $T_\mathrm{in} = T_*$ because the free-streaming length until $T=T_*$ is negligible.
The velocity $v(t)$ is given by 
\begin{align}
    v  
    &=
    \frac{p}{\sqrt{p^2+m_s^2}}
    =
    \frac{\epsilon_{\rm{phys}}(T_{*})}
    {\sqrt{ \epsilon_{\rm{phys}}(T_{*})^2+(\frac{m_s}{T_{*}})^2(\frac{a}{a_*})^2 }}
    \ ,
    \label{eq : velocity and momentum}
\end{align}
where we used $p = \epsilon_{\rm{phys}}(T_{*}) T_{*} a_*/a$.
Now, we calculate the free-streaming length of the sterile neutrinos in the SF mechanism using the distribution function obtained in Sec.~\Ref{sec: calculation of Shi Fuller mechanism}.
We show the free-streaming length of the sterile neutrinos as a function of $m_s$ for a given $L_{\nu_e}^{\rm{init}}$ in Fig.~\ref{fig : Free streaming length}. 
By equating $\lambda_{\rm{FS}}(m_s,L_{\nu_e}^{\rm{init}})$ and $\lambda_{\rm{FS}}^{\rm{upper}}=0.099$\,Mpc, we obtain the Lyman-$\alpha$ constraint on $m_s$ for a given $L_{\nu_e}^{\rm{init}}$. 
We show the lower bounds of $m_s$ for several values of $L_{\nu_e}^{\rm{init}}$ in Table~\ref{Table : Lyman alpha constraint on sterile neutrino mass}.
\begin{table}[htbp]
    \renewcommand{\arraystretch}{1.1}
    \centering
    \begin{tabular}{|c||c|c|c|c|c|c|c|}
        \hline
        $L_{\nu_e}^{\rm{init}}$  & 0 (DW) & $10^{-5}$ &$10^{-4.5}$ &$10^{-4}$ & $10^{-3.5}$ & $10^{-3}$ & $5\times 10^{-3}$
        \\
        \hline\hline
        $m_s^{\rm{lower}}$ [keV] &  21 & 16& 11 & 18 & 22 & 23 & 24
        \\
        \hline
    \end{tabular}
    \caption{Lyman-$\alpha$ constraints on sterile neutrino mass for given values of $L_{\nu_e}^{\rm{init}}$.}
    \label{Table : Lyman alpha constraint on sterile neutrino mass}
\end{table}

The constraint on $m_s$ becomes weaker for $L_{\nu_e}^{\rm{init}}\lesssim 10^{-4.5}$ and stronger for $L_{\nu_e}^{\rm{init}}\gtrsim 10^{-4.5}$ as $L_\mathrm{\nu_e}^\mathrm{init}$ increases. 
This trend can be understood as a combination of two effects.
First, as discussed in Sec.~\ref{subsec : analytical study}, the averaged momentum of resonantly produced sterile neutrinos becomes smaller than the thermal distribution at production.
In particular, this effect is significant if the substantial fraction of $L_{\nu_e}^\mathrm{init}$ is consumed in the resonant production.
Second, the resonance temperature depends on lepton asymmetry as $T_\mathrm{low} \propto L_{\nu_e}^{-1/4}$ (see Eq.~\eqref{input parameter dependence of resonance condition}).
Thus, larger $L_{\nu_e}^\mathrm{init}$ results in later production of sterile neutrinos, which are less red-shifted.
Consequently, in spite of the first effect, the averaged momentum of sterile neutrinos for large $L_{\nu_e}^\mathrm{init}$ can be larger than that in the DW mechanism when compared at the same temperature.
Due to the balance between these effects, the Lyman-$\alpha$ constraint becomes weakest for $L_{\nu_e}^\mathrm{init} \simeq 10^{-4.5}$.

\begin{figure}[t]
\centering
\includegraphics[width=0.7\linewidth]{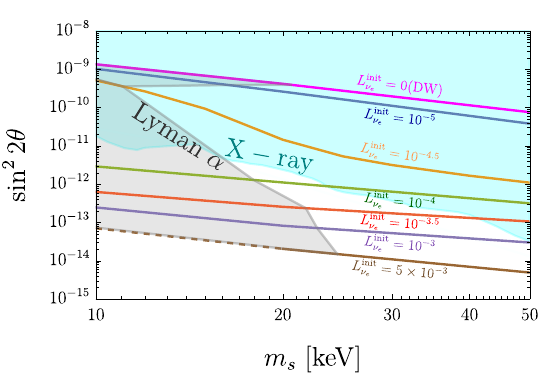}
    \caption{%
    $m_s$ and $\sin^2 2\theta$ to explain all dark matter with the Lyman-$\alpha$ constraint (gray) and the X-ray constraint (light blue).
    We adopted the conservative X-ray constraint from Ref.~\cite{Boyarsky:2018tvu}, which is based on  
    the $\textit{NuStar}$ observations~\cite{Neronov:2016wdd}. 
    }
    \label{fig : Contour plot including Lyman alpha}
\end{figure}

We show the $m_s$\,--\,$\sin^2 2\theta$ contour to explain all dark matter and observational constraints in Fig.~\Ref{fig : Contour plot including Lyman alpha}.
We see that $L_{\nu_e}^{\rm{init}}\gtrsim \mathcal{O}(10^{-4})$ is required to evade the current X-ray constraint, and $m_s\gtrsim 20$\,keV is required to evade the current Lyman-$\alpha$ constraint. 

\subsection{Gravitational wave enhancement at L-ball decay}
\label{subsec: L-ball decay GW}

In our scenario, the energy density of L-balls dominates the universe before the L-ball decay.
Then, at the L-ball decay, the universe undergoes a sudden transition from the early matter-dominated (eMD) era to the radiation-dominated (RD) era.
During the eMD era, the subhorizon mode of the gravitational potential is almost constant and does not decay in contrast to in the RD era.
Then, at the sudden transition to the RD era, the gravitational potential begins to oscillate due to radiation pressure.
This process induces a large value of the fluid velocity and enhances the scalar-induced gravitational waves.
This mechanism of an enhancement of scalar-induced gravitational waves is called ``Poltergeist'' mechanism~\cite{Inomata:2020tkl} and has been studied in the context of the L-ball decay~\cite{Kasuya:2022cko,Kawasaki:2023rfx,White:2021hwi}.
Here, we evaluate the gravitational wave spectrum in our scenario and discuss the testability in future observations. 

In the following, we introduce a conformal time $\eta \equiv \int{\rm{d}}t/a(t)$.
We denote $\eta$ at the completion of reheating by $\eta_{\rm{R}}$, $\eta$ when the energy density of L-balls begins to dominate the universe by $\eta_{\rm{eq,1}}$, $\eta$ when the gravitational potential $\Phi$ begins to decouple from the matter density perturbation $\delta_{\rm{m}}$ by $\eta_{\rm{eq,2}}$, and $\eta$ at the L-ball decay by $\eta_{\rm{decay}}$.

The approximate analytic study gives the density parameter of gravitational waves per $\ln k$, $\Omega_{\rm{GW}}$,  as~\cite{Inomata:2020tkl,Kawasaki:2023rfx}%
\footnote{%
The production of gravitational waves by this mechanism is also studied numerically in Ref.~\cite{Pearce:2023kxp}.
}
\begin{align}
    \Omega_{\rm{GW}}(\eta_c,k)
    \simeq
    2.9\times 10^{-7}Y\mathcal{P}_{\zeta}^2\Phi_{\rm{low}}^4(k)(k\eta_{\rm{decay}})^7F(n_{\rm{s,eff}})
    \ ,
    \label{eq : approximate analytic formula of GW spectrum}
\end{align}
where $\Omega_\mathrm{GW}$ is defined by 
\begin{align}
    \Omega_{\rm{GW}}(\eta,k)
    &\equiv
    \frac{\rho_{\rm{GW}}(\eta,k)}{\rho_{\rm{tot}}(\eta)} 
    \nonumber \\
    &=
    \frac{1}{24}\left(\frac{k}{\mathcal{H}(\eta)}\right)^2\overline{\mathcal{P}_{h}(\eta,k)}
    \ .
\end{align}
For details on this formula, see Appendix~\Ref{Appendix : GW from Lball decay}. Here, $\rho_{\rm{tot}}(\eta)$ is the total energy density at $\eta$, $\rho_{\rm{GW}}(\eta,k)$ is the energy density of gravitational waves per $\ln k$, $\overline{\mathcal{P}_h(\eta,k)}$ is the time average of the gravitational wave spectrum, and $\mathcal{H}\equiv aH$ is the conformal Hubble parameter. 
We evaluated the spectrum at a certain time $\eta=\eta_c$ after the production of the gravitational waves and before the late time matter-radiation equality. 
In Eq.~\eqref{eq : approximate analytic formula of GW spectrum}, $Y\simeq 2.3$ is a numerical fudge factor, and $F(n_{\rm{s,eff}})$ is defined using a hypergeometric function ${}_2F_1$ as
\begin{align}
    F(n_{\rm{s,eff}})
    \equiv &
    2\left(\frac{3}{4}\right)^{n_{\rm{s,eff}}-1}
    \nonumber \\
    &\times 
    \left[
        4 {}_2F_1\left(\frac{1}{2},1-n_{\rm{s,eff}};\frac{3}{2};\frac{1}{3}\right)
        -3{}_2F_1\left(\frac{1}{2},-n_{\rm{s,eff}};\frac{3}{2};\frac{1}{3}\right)
        -{}_2F_1\left(\frac{3}{2},-n_{\rm{s,eff}};\frac{5}{2};\frac{1}{3}\right)
    \right]
    \ ,
\end{align}
where $n_{\rm{s,eff}}$ represents the effective spectral index of $\Phi_{\rm{low}}$, defined by
\begin{align}
    n_{\rm{s,eff}}
    \equiv
    1+2\frac{{\rm{d}}\ln\Phi_{\rm{low}}}{{\rm{d}}\ln k}
    \ .
\end{align}
Here, $\Phi_{\rm{low}}$ is the lower bound of $\Phi$ at the onset of RD era and is given by Eq.~\eqref{Phi low}.
$\mathcal{P}_\zeta$ is the power spectrum of the curvature perturbations, and we assume that it is given by the scale-invariant spectrum as
\begin{align}
    \mathcal{P}_\zeta (k)
    =
    C^2 A_\mathrm{s} 
    \ ,
\end{align}
where $A_\mathrm{s} = 2.1 \times 10^{-9}$ is the amplitude on the CMB pivot scale $k_* = 0.05~\mathrm{Mpc}^{-1}$~\cite{Planck:2018vyg}, and a constant $C$ is introduced because the amplitude can be much larger at small scales.
We translate $\Omega_\mathrm{GW}(\eta_c,k)$ into the current value of the gravitational wave energy density parameter as
\begin{align}
    \Omega_{\rm{GW},0}h^2
    =
    0.83\left(\frac{g_{*,c}}{10.75}\right)^{-1/3}\Omega_{\rm{r},0}h^2\Omega_{\rm{GW}}(\eta_c,k)
    \ ,
\end{align}
which depends on three parameters: $C$, $\eta_{\rm{eq,1}}$, and $\eta_{\rm{decay}}$.
In other words, the gravitational wave spectrum depends on the amplitude of the scalar perturbations, the decay time of the L-balls, and the duration of the eMD era.

To quantify $\eta_{\rm{eq,1}}/\eta_{\rm{decay}}$ as a function of the parameters on the L-ball scenario, we use the relation between the scale factor and $\eta$ with $\eta_{\rm{eq,1}}\lesssim \eta\lesssim \eta_{\rm{decay}}$:
\begin{align}
    \frac{a(\eta)}{a(\eta_{\rm{eq,1}})}
    =
    \left(
        \frac{\eta}{\eta_*}
    \right)^2
    +\frac{2\eta}{\eta_*} 
    \simeq
    \left(
        \frac{\eta}{\eta_*}
    \right)^2
    \ ,
    \label{determining a/aeq1 0}
\end{align}
where $\eta_*\equiv \eta_{\rm{eq,1}}/(\sqrt{2}-1)$.
In the last equality, we assumed $\frac{a(\eta)}{a(\eta_{\rm{eq,1}})}\gg 1$.
Using Eqs.~\eqref{Tdecay}, \eqref{determining a/aeq1 4}, and \eqref{determining a/aeq1 0}, we obtain $\eta_{\rm{decay}}/\eta_{\rm{eq,1}}$ as a function of the basic parameters, $m_{3/2}, M_{\rm{F}}, T_{\rm{R}}$, and $\varphi_{\rm{osc}}$. 

\begin{table}[b]
    \renewcommand{\arraystretch}{1.1}
    \centering
    \begin{tabular}{|c|c|c|c|c||c|c|}
    \hline
    & $m_{3/2}$ [GeV] & $M_{\rm{F}}$ [GeV] & $T_{\rm{R}}$ [GeV] & $M_s$ [GeV] &$|\eta_{\rm{b}}| $&$|L_{\nu_e}^{\rm{init}}|$
    \\
    \hline\hline
    A & 0.5 & $1.1\times10^{6}$ & $1.0\times10^2$ & $1.5\times10^4$ &$8.7\times10^{-11}$&$7.0\times 10^{-4}$  
    \\
    \hline
    B & 0.4 & $8.5\times10^{5}$ & $1.0\times10^3$ & $1.7\times10^5$ & $8.7\times10^{-11} $&$8.4\times 10^{-4}$  
    \\
    \hline
    C & 2.0 & $5.4\times10^{6}$ & $1.4\times 10^4$ & $2.1\times10^5$ & $8.7\times10^{-11} $&$2.2\times 10^{-4}$  
    \\
    \hline
    \end{tabular}
    \caption{Benchmark parameters to evaluate the gravitational wave spectra. 
    }
    \label{Table : benchmark parameters}
\end{table}
\begin{figure}[t]
    \centering
    \includegraphics[width=1.0\linewidth]{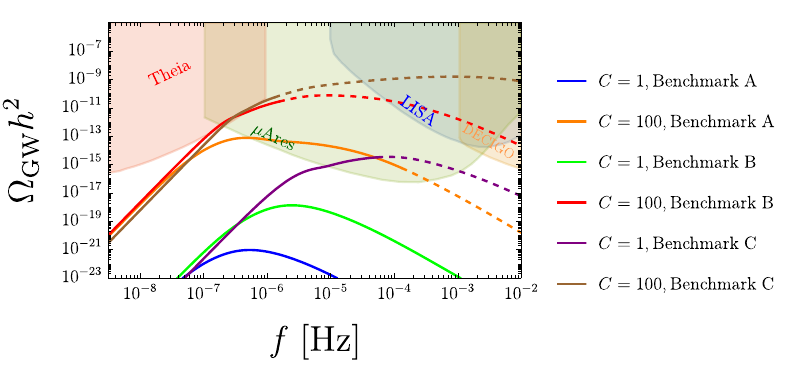}
    \caption{%
    Gravitational wave spectra enhanced at the L-ball decay.
    The analytic formula becomes unreliable in the dashed regions because the fitting formula~\eqref{eq : fNL} of non-linear density perturbations in the eMD era may not be applicable in the wavelength corresponding to these regions.
    }
    \label{fig : Second order GW}
\end{figure}
We evaluated the final gravitational wave spectrum with three benchmark parameters shown in Table.~\Ref{Table : benchmark parameters}. 
In these parameter sets, we fixed $\varphi_{\rm{osc}}/M_{\rm{Pl}}=0.3$ and $|\delta|=10^{-3}$.
We show the results in Fig.~\Ref{fig : Second order GW}.
We see that the gravitational waves in cases A and B with $C=100$, and C with $C=1,100$ are detectable in future gravitational wave observations such as $\mu$Ares~\cite{Sesana:2019vho} and THEIA~\cite{Theia:2019non}.
Moreover, the gravitational wave abundance increases in the order of cases C, B, and A for the same value of $C$. 
This is because of the difference in $T_{\rm{R}}$. 
If the reheating temperature is low and near the decay temperature of L-balls (case A), the L-ball domination era lasts for a short period, and the gravitational wave spectrum is not enhanced due to the suppression of subhorizon modes of the scalar perturbations during the eRD era.
On the other hand, for $T_{\rm{R}}\sim \mathcal{O}(10^4\,\text{--}\,10^5)\,{\rm{GeV}}\gg T_{\rm{decay}}$, the eMD era lasts long, and the gravitational wave spectrum is significantly enhanced. 
Note also that a large gaugino mass $M_s \gtrsim \mathcal{O}(10^5)$\,GeV is assumed in cases B and C.
This is because the baryon asymmetry is overproduced due to a large evaporation rate of L-balls for a higher value of $T_{\rm{R}}$,
unless we assume a large gaugino mass.
Note that we fixed $M_s$ in each case so that the baryon asymmetry of the universe $\eta_b^{\rm{obs}} \sim 10^{-10}$ can be explained by the L-ball evaporation process and sphaleron process above the electroweak scale. 

Finally, we note that our scenario assumes larger values of $T_{\rm{decay}}$ than Ref.~\cite{Kawasaki:2023rfx}, to realize the SF mechanism. 
Consequently, we obtain enhanced gravitational waves in a higher frequency region compared with Ref.~\cite{Kawasaki:2023rfx}.

\section{Conclusion}
\label{sec: conclusion}

In this paper, we have investigated the possibility that a large lepton asymmetry generated from L-ball decay induces resonant production of sterile neutrino dark matter.
We revisited the numerical calculation of the SF mechanism.
As a result, we found that the value of the vacuum mixing angle $\theta$ to explain all dark matter largely deviates from the DW case for $L_{\nu_e}^{\rm{init}} > \mathcal{O}(10^{-4})$ in the case of $m_s>10$\,keV.

We then discussed the AD leptogenesis scenario with the L-ball decay, which produces a large lepton asymmetry in the active neutrino sector as a source of sterile neutrino production.
Considering the delayed-type L-balls, we found consistent parameter regions for which sterile neutrinos are resonantly produced through the SF mechanism and account for all the dark matter of the universe.

Finally, we discussed the observational implications of our scenario.
We evaluated the free-streaming length of the produced sterile neutrinos from the distribution functions obtained in the numerical calculation.
As a result, it was found that the sterile neutrino mass should be $m_s \gtrsim 10$\,\text{--}\,$20$\,keV to be consistent with the current Lyman-$\alpha$ constraint.
We also found that the constraint on $m_s$ becomes weaker with larger $L_{\nu_e}^{\rm{init}}$ for $L_{\nu_e}^{\rm{init}}\lesssim 10^{-4.5}$ while the constraint becomes stronger with larger $L_{\nu_e}^{\rm{init}}$ for $L_{\nu_e}^{\rm{init}}\gtrsim 10^{-4.5}$. 
We also discussed the scalar-induced gravitational waves enhanced at the L-ball decay.
We found that our scenario is testable in future gravitational wave observations such as $\mu$Ares and THEIA. 
However, to discuss the testability of higher frequency modes, $f\gtrsim \mathcal{O}(10^{-5})$\,Hz, where LISA and DECIGO have good sensitivity, further discussion on the effect of non-linear density evolution on the gravitational potential during the Q-ball dominated era is required.

\begin{acknowledgments}
This work was supported by JSPS KAKENHI Grant Nos. 20H05851(M.K.), 21K03567(M.K.), 23KJ0088 (K.M.), and JST SPRING (grant number: JPMJSP2108) (K.K.).
K.K. was supported by the Spring GX program.
\end{acknowledgments}

\appendix

\section{Fitting function for \texorpdfstring{$y_e$}{}}
\label{Appendix : formula}

\begin{figure}[t]
\centering
    \includegraphics[width=0.8\linewidth]{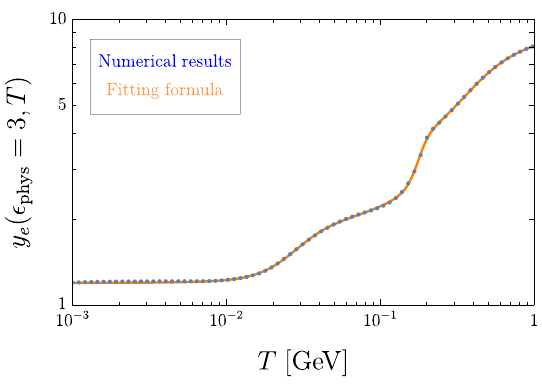}
    \caption{Fitting formula~\eqref{fitting formula of thermal width} for $y_e$ used in this paper (orange line) and the numerical results presented in Ref.~\cite{Asaka:2006nq_web} (blue dots).
    }
    \label{fig : comparison of fitting formula of ye with text file data}
\end{figure}

To numerically solve the master equation~\eqref{master equation of nus distribution}, we should take into account the temperature dependence of $y_e(\epsilon,T)$. 
In this paper, we use a function fitted to the numerical result given in Refs.~\cite{Asaka:2006nq,Asaka:2006nq_web}. 
For simplicity, we approximate that $y_e$ does not depend on $\epsilon_{\rm{phys}}$ and use the data for $\epsilon_{\rm{phys}}(T)=3$ in Ref.~\cite{Asaka:2006nq_web}.
The fitting formula is given by
\begin{equation}
    y_e(\epsilon, T)= \sum_{i=1}^{4}\exp\left[c_1^i+c_2^i\left(1+\tanh\left(\frac{\log \left(\frac{T}{1\,{\rm{GeV}}}\right)+c_3^i}{c_4^i}\right)\right)\right],
    \label{fitting formula of thermal width}
\end{equation}
where the values of $c_i^j$ are shown in Table~\ref{Table : Fitting parameters of ye}.
We compare the formula~\eqref{fitting formula of thermal width} with the numerical result in Ref.~\cite{Asaka:2006nq_web} in Fig.~\ref{fig : comparison of fitting formula of ye with text file data}.

We note that the value of $y_e$ does not affect the final result in the limit where the time evolution of the lepton asymmetry is negligible during the production of a single mode as we can see that Eq.~\eqref{analytic result of contribution from low resonance point} is independent of $y_e$. 
On the other hand, if the lepton asymmetry non-negligibly evolves during the resonance of a single mode with fixed $\epsilon$, $y_e$ slightly affects the final distribution of the sterile neutrino.
\begin{table}[tb]
    \renewcommand{\arraystretch}{1.1}
    \centering
    \begin{tabular}{|c|c|c|c|c|} 
    \hline
    $i$ & $c_1^i$ & $c_2^i$ & $c_3^i$ & $c_4^i$
    \\
    \hline\hline
    $1$ & $-0.848259$ & $0.685952$ & $1.76853$ & $0.174619$ 
    \\
    \hline
    $2$ & $ -2.36415$ & $ 1.55878$ & $ 1.2981$ & $0.656795$ 
    \\
    \hline
    $3$ & $ -1.52151$ & $ 1.50891$ & $1.64656$ & $  1.5446$ 
    \\
    \hline
    $4$ & $-0.792215$ & $0.464922$ & $3.63313$ & $0.572627$ 
    \\
    \hline
    \end{tabular}
    \caption{Fitting parameters for $y_e$, $c_j^i\;(i, j=1, \cdots, 4$).}
    \label{Table : Fitting parameters of ye}
\end{table}

\section{Numerical dependence on the number of bins used in momentum space}
\label{Appendix : bin dependence}

To solve Eqs.~\eqref{master equation of nus distribution} and~\eqref{eq : time evolution of lepton asymmetry} simultaneously, we discretize the momentum space as $\epsilon=\epsilon_{\rm{max}}/N_{\rm{bin}}\times i$, where $N_{\rm{bin}}$ is the number of momentum bins, $i = 1, \cdots, N_\mathrm{bin}$, and $\epsilon_{\rm{max}}$ is the maximum value of $\epsilon$ in the numerical calculation.
Here, we fix $\epsilon_{\rm{max}}=20$ and observe the dependence of the final abundance $\Omega_{\nu_s}$ on $N_{\rm{bin}}$.

\begin{figure}[t]
\centering
        \includegraphics[width=0.8\linewidth]{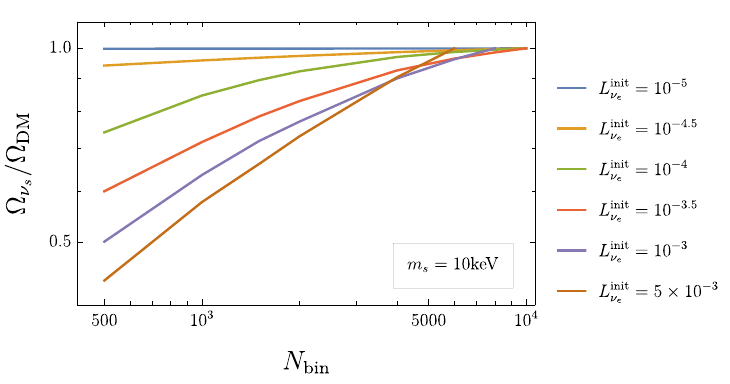}
        \\
        \includegraphics[width=0.8\linewidth]{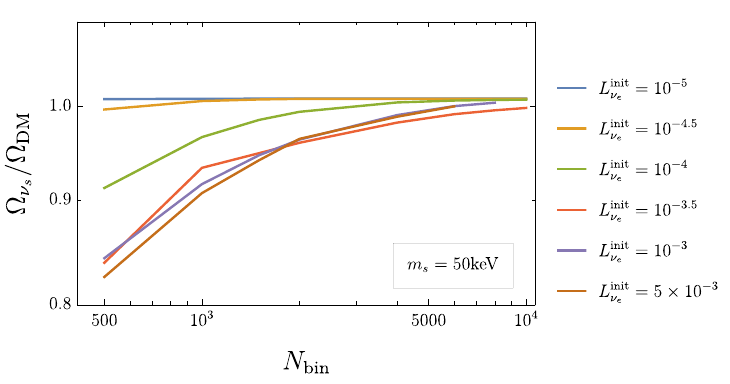}
    \caption{Dependence of the sterile neutrino abundance $\Omega_{\nu_s}$ on $N_{\rm{bin}}$ with fixed $\theta$.
    }
    \label{fig : dependence of final abundance on Nbin}
\end{figure}

We show the dependence of the sterile neutrino abundance $\Omega_{\nu_s}$ on $N_{\rm{bin}}$ with fixed $\theta$ in Fig.~\ref{fig : dependence of final abundance on Nbin}. 
Here, we chose $\theta$ to realize $|\Omega_{\nu_s}/\Omega_{\rm{DM}}-1| < \mathcal{O}(10^{-3})$ with the value of $N_{\rm{bin}}$ used in Sec.~\ref{sec: calculation of Shi Fuller mechanism} ($=N_{\rm{bin}}^{\rm{max}}$).
We summarize the values of $N_{\rm{bin}}^{\rm{max}}$ in Table.~\ref{Table : maximum Nbin}. 
Note that $N_{\rm{bin}}^{\rm{max}}$ does not depend on $m_s$.
These are the maximum values due to the limitation of the performance of our computer memories.
From Fig.~\ref{fig : dependence of final abundance on Nbin}, we see that the dependence on $N_{\rm{bin}}$ is insignificant at $N_\mathrm{bin} = N_\mathrm{bin}^\mathrm{max}$ with $L_{\nu_e}^{\rm{init}}\lesssim 10^{-4}$. On the other hand, with $L_{\nu_e} \gtrsim 10^{-3.5}$, $\Omega_{\rm{\nu_s}}$ still increases at $N_\mathrm{bin} = N_\mathrm{bin}^\mathrm{max}$.
\begin{table}[b]
    \renewcommand{\arraystretch}{1.1}
    \centering
    \begin{tabular}{|c||c|c|c|c|c|c|} 
        \hline
        $L_{\nu_e}^{\rm{init}}$  &  $10^{-5}$ &$10^{-4.5}$ &$10^{-4}$ & $10^{-3.5}$ & $10^{-3}$ & $5\times 10^{-3}$
        \\
        \hline\hline
        $N_{\rm{bin}}^{\rm{max}}$  &   $10^4$& $10^4$ & $10^4$ & $10^4$& 8000 & 6000 
        \\
        \hline
    \end{tabular}
    \caption{$N_{\rm{bin}}^{\rm{max}}$ used in the numerical studies in Sec.~\Ref{sec: calculation of Shi Fuller mechanism}.
    }
    \label{Table : maximum Nbin}
\end{table}

The reason can be qualitatively understood as follows. 
If $L_{\nu_e}$ decreases significantly during a resonant production of a single mode $\epsilon$, the resonance terminates earlier because $T_\mathrm{low}$ grows as $T_\mathrm{low} \propto L_{\nu_e}^{-1/4}$.
If we use large $\Delta \epsilon$, the decrease of $L_{\nu_e}$ during the resonance of a given $\epsilon$ is overestimated, and thus the resonant production of sterile neutrinos is underestimated.
Thus, the estimation with constant $L_{\nu_e}$ gives an upper bound on the final abundance. For instance, with $m_s=10$\,keV and $L_{\nu_e}^{\rm{init}}=5\times 10^{-3}$ (the case in which the final spectrum dependence on $N_{\rm{bin}}$ is the strongest), the final abundance becomes $\Omega_{\nu_s}/\Omega_{\rm{DM}}=1.46$ with the same $\theta$ as in the contour in Fig.~\Ref{fig : Contour Plot}, assuming that $L_{\nu_e}$ is constant. Thus, we conclude that at least our estimation has an accuracy of several tens of percent.

\section{Analytical estimate of the sterile neutrino spectrum}
\label{Appendix : approximate integration of master equation}

Here, we show the derivation of the analytical estimate of the sterile neutrino spectrum in Sec.~\ref{subsec : analytical study}.
If the sterile neutrinos are produced mainly via the SF mechanism, the contribution of anti-neutrinos is negligible, and the time evolution of $f_{\nu_s}$ is described by
\begin{equation}
    \frac{{\rm{d}}}{{\rm{d}}T} f_{\nu_s}(\epsilon,T)
    \simeq
    \frac{{\rm{d}}t}{{\rm{d}}T}
   \Gamma_{\nu_a}(\epsilon,T)
        \theta_M^2(\epsilon,T) f_{\nu_e}(\epsilon,T)
    \ .
    \label{eq : integrant}
\end{equation}
Now, we integrate the right-hand side over $T$.
Since most of the sterile neutrinos are produced around $T=T_{\rm{low}}$, we focus on  $T \sim T_{\rm{low}}$.
Then, $\theta_M$ is approximately given as 
\begin{align}
\theta_M^2
    &=
    \theta^2
    \left[
        \left(1 - \frac{2p}{m_s^2}V_{a}(\epsilon,T) \right)^2
        + \frac{p^2\Gamma_{\nu_a}^2(\epsilon,T)}{m_s^4}
    \right]^{-1}
    \nonumber\\
   &\simeq
    \theta^2
    \left[
    x(\epsilon,T)
    ^2
    +
    y(\epsilon,T)^2
    \right]^{-1}
    \ ,
\end{align}
where 
\begin{align}
    x(\epsilon,T)
    &\equiv
    1 - \frac{8\sqrt{2}\pi^2 G_{\rm{F}} \epsilon_{\rm{phys}}(T) L_{\nu_e} g_{*}(T)T^4}{45 m_s^2}
    \ ,
    \\ 
    y(\epsilon,T) 
    &\equiv
    \frac{\epsilon_{\rm{phys}}(T) \Gamma_{\nu_a}(\epsilon,T) T}{m_s^2}
    \ .
\end{align}
Here, we assume that the time evolution of $L_{\nu_e}$ is negligible, and then $x$ is a monotonically decreasing function of $T$.
As a result, we can integrate Eq.~\eqref{eq : integrant} with respect to $x$ and obtain
\begin{align}
    f_{\nu_s}
    \simeq 
    \int_{x_{\rm{min}}}^{x_{\rm{max}}}
    {\rm{d}}x \,
    \left|
        \frac{{\rm{d}}T}{{\rm{d}}x}
        \frac{{\rm{d}}t}{{\rm{d}}T}
    \right|
    \frac{\theta^2 f_{\nu_e}(\epsilon,T)) \Gamma_{\nu_e}(\epsilon,T)}
    {x^2+y(\epsilon,T)^2}
    \ ,
    \label{eq : f integrated wrt x}
\end{align}
where $x_{\rm{min}}$ and $x_{\rm{max}}$ denote the integration range of Eq.~\eqref{eq : integrant}.
Since the resonance occurs at $x \simeq 0$, the contribution to $f_{\nu_s}$ mainly comes from the integration around $x = 0$, which corresponds to $T = T_\mathrm{low}$.
Thus, we neglect the temperature dependence of the integrand except for the resonant part and obtain
\begin{align}
    f_{\nu_s} 
    \simeq 
    \left|
        \left.
            \frac{{\rm{d}}T}{{\rm{d}}x}
        \right|_{T_{\rm{low}}}
        \left.
            \frac{{\rm{d}}t}{{\rm{d}}T}
        \right|_{T_{\rm{low}}}
    \right|
    \theta^2
    \Gamma_{\nu_e}(\epsilon,T_{\rm{low}})
    f_{\nu_e}(\epsilon,T_{\rm{low}})
    \int_{x_{\rm{min}}}^{x_{\rm{max}}}
    \frac{{\rm{d}}x}{x^2+y(\epsilon,T_{\rm{low}})^2}
    \ .
    \label{eq : approximation of integrant}
\end{align}
Due to the resonant feature of the integrand, we can approximately set $x_{\rm{min}}=-\infty$ and $x_{\rm{max}}=\infty$.
Consequently, we obtain
\begin{align}
    f_{\nu_s} 
    &\simeq 
    \left|
        \left.
            \frac{{\rm{d}}T}{{\rm{d}}x}
        \right|_{T_{\rm{low}}}
        \left.
            \frac{{\rm{d}}t}{{\rm{d}}T}
        \right|_{T_{\rm{low}}}
    \right|
    \theta^2
    \Gamma_{\nu_e}(\epsilon,T_{\rm{low}})
    f_{\nu_e}(\epsilon,T_{\rm{low}})
    \frac{\pi}{y(\epsilon,T_{\rm{low}})}
    \nonumber \\
    &=
    \frac{135\sqrt{5}}{32 \pi^2}
    \frac{g_*^{2/3}(T_i)}{g_*^{13/6}(T_{\rm{low}})}
    \frac{M_{\rm{Pl}} m_s^4 \theta^2}{G_{\rm{F}} \epsilon^2 T_{\rm{low}}^7}
    \frac{f_{\nu_e}(\epsilon,T_{\rm{low}})}{L_{\nu_e}(T_{\rm{low}})}
    \ .
    \label{eq : approximate fnus low}
\end{align}
In a similar way, we also obtain the contribution from the resonance at $T_{\rm{high}}$ as
\begin{align}
    f_{\nu_s} 
    &\simeq
    \frac{135\sqrt{5}}{16 \pi^2}
    \frac{g_*^{2/3}(T_i)}{g_*^{13/6}(T_{\rm{high}})}
    \frac{1+ \frac{1}{3}\frac{{\rm{d}}\ln g_{*}}{{\rm{d}}\ln T}(T_{\rm{high}})}
    {1 - \frac{1}{3}\frac{{\rm{d}}\ln g_{*}}{{\rm{d}}\ln T}(T_{\rm{high}})}
    \frac{M_{\rm{Pl}} m_s^4 \theta^2}{G_{\rm{F}} \epsilon^2 T_{\rm{high}}^7}
    \frac{f_{\nu_e}(\epsilon,T_{\rm{high}})}{L_{\nu_e}(T_{\rm{high}})}
    \ .
    \label{eq : approximate fnus high}
\end{align}

Note that the factor $(1+ \frac{1}{3}\frac{{\rm{d}}\ln g_{*}}{{\rm{d}}\ln T}(T_{\rm{high}}))/
    (1 - \frac{1}{3}\frac{{\rm{d}}\ln g_{*}}{{\rm{d}}\ln T}(T_{\rm{high}}))\simeq 1$ is omitted in Eq.~\eqref{analytic result of contribution from high resonance point}.
This approximation becomes less accurate for a smaller value of $m_s$, fixing the final abundance of sterile neutrinos. 
This is because we cannot regard $L_{\nu_e}$ as constant during the resonance of a single mode $\epsilon$. 

\section{Review of gravitational waves enhanced at the L-ball decay}
\label{Appendix : GW from Lball decay}

In this appendix, we briefly review the basic formalism to evaluate the gravitational wave spectrum enhanced at the L-ball decay based on Refs.~\cite{Inomata:2020tkl,Kasuya:2022cko,Kawasaki:2023rfx}.
Here, we use the conformal Newtonian gauge.
Then, the metric perturbation is given by
\begin{equation}
    {\rm{d}}s^2
    =
    a^2
    \left[
    -(1+2\Phi){\rm{d}}\eta^2
    +
    \left(
    (1-2\Psi)\delta_{ij}+\frac{1}{2}h_{ij}
    \right)
    {\rm{d}}x^i{\rm{d}}x^j
    \right],
\end{equation}
where $\Phi$ and $\Psi$ is the first-order scalar perturbations, and $h_{ij}$ is the tensor perturbation.
In the following, we neglect the anisotropic stress and take $\Phi=\Psi$. 

Now, we discuss the second-order tensor perturbation $h_{ij}$ induced by the first-order scalar perturbation $\Phi$.
The time-averaged gravitational wave spectrum $\overline{\mathcal{P}_h(\eta,k)}$ induced by the first-order scalar perturbations is given by~\cite{Kohri:2018awv,Ananda:2006af,Baumann:2007zm}
\begin{equation}
    \overline{\mathcal{P}_h(\eta,k)}
    =
    4\int_0^\infty {\rm{d}}v
    \int_{|1-v|}^{1+v}{\rm{d}}u 
    \left(
    \frac{4v^2-(1+v^2-u^2)^2}{4uv}
    \right)^2
    \overline{I^2(u,v,k,\eta_{\rm{R}},\eta)}
    \mathcal{P}_\zeta(uk)
    \mathcal{P}_\zeta(vk)
    \ ,
    \label{eq : relation between GW spectrum and curvature spectrum}
\end{equation}
where $\mathcal{P}_\zeta$ is the primordial spectrum of the curvature perturbations. 
Here, $\overline{I^2(u,v,k,\eta_{\rm{R}},\eta)}$ is the time average of $I^2$ defined by
\begin{equation}
   I(u,v,k,\eta_{\rm{R}},\eta)
    \equiv 
    \int_0^x {\rm{d}}\tilde{x}
    \frac{a(\tilde{\eta})}{a(\eta)}
    kG_k(\eta,\tilde{\eta})
    f(u,v,\tilde{x},x_{\rm{decay}})
    \ ,
    \label{eq : I}
\end{equation}
where $x\equiv k\eta$.
Here, $G_k(\eta,\tilde{\eta})$ is Green's function satisfying 
\begin{equation}
    G_k''(\eta,\tilde{\eta})+
    \left(
        k^2-\frac{a''(\eta)}{a(\eta)}
    \right)
    G_k(\eta,\tilde{\eta})
    =
    \delta(\eta-\tilde{\eta})
    \ ,
\end{equation}
and $f$ is defined by
\begin{align}
    &f(u,v,\tilde{x},x_{\rm{decay}})
    \nonumber\\
    &\equiv
    \frac
    {3\left[
        2(5+3w(\tilde{\eta}))
        \mathcal{T}(u\tilde{x})\mathcal{T}(v\tilde{x})
        +4\mathcal{H}^{-1}
        (\mathcal{T}(u\tilde{x})\mathcal{T}'(v\tilde{x})
        +\mathcal{T}'(u\tilde{x})\mathcal{T}(v\tilde{x}))
        +4\mathcal{H}^{-2}
        \mathcal{T}'(u\tilde{x})\mathcal{T}'(v\tilde{x})
    \right]}
    {25(1+w(\tilde{\eta}))}
    \ ,
    \label{eq : f}
\end{align}
where $w(\tilde{\eta})$ is the equation-of-state parameter defined by $p=w \rho$ at $\eta=\tilde{\eta}$, and $\mathcal{H}\equiv aH$ is the conformal Hubble parameter.
Here, $\mathcal{T}(k\eta)$ is the transfer function of scalar perturbation $\Phi$, defined by $\Phi_{\mathbf{k}}\equiv\phi_{\mathbf{k}}\mathcal{T}(k\eta)$, where $\Phi_{\mathbf{k}}$ is the Fourier component of $\Phi$, and $\phi_{\mathbf{k}}$ is the primordial value of $\Phi_{\mathbf{k}}$.

To calculate $\overline{\mathcal{P}_h(\eta,k)}$, we consider the time evolution of $\mathcal{T}(x)$ by dividing the evolution into four eras:
    (i) eRD era,
    (ii) eMD era,
    (iii) transition period to RD era during L-ball decay process,
    and (iv) RD era.
Firstly, as for era (i), $\Phi(x)$ for superhorizon modes remains almost constant while $\Phi(x)$ for subhorizon modes is suppressed.
The transfer function at $\eta=\eta_{\rm{eq,1}}$ is evaluated by a fitting formula~\cite{Bardeen:1985tr,Inomata:2020lmk}
\begin{align}
   &\mathcal{T}(x_{\rm{eq,1}})
   \nonumber \\
   &=
   \frac{\ln\left[1+0.146x_{\rm{eq,1}}\right]}{0.146x_{\rm{eq,1}}}
   \left[1+0.242x_{\rm{eq,1}}+(1.01x_{\rm{eq,1}})^2
   +(0.341x_{\rm{eq,1}})^3+(0.418x_{\rm{eq,1}})^4\right]^{-1/4}
   \ .
   \label{transfar function at eRD end}
\end{align}

Secondly, as for era (ii), $\mathcal{T}$ remain constant in the linear theory regime, and density perturbations $\delta_{\rm{m}}$ grow as $\propto a$.
Then, $\delta_{\rm{m}}$ with a certain mode enters the non-linear regime during the eMD era, and we need
to take into account non-linear effects. 
We assume that the density perturbation in the non-linear regime $\delta_{\rm{m}}(k_{\rm{NL}})$ and gravitational potential is connected by Poisson equation given by 
\begin{align}
   \frac{3}{5}k_{\rm{NL}}^2\mathcal{T}(x_{\rm{NL}})
   \mathcal{P}_\zeta^{1/2}
   =
   \frac{3}{2}\mathcal{H}^2
   \delta_{\rm{m,NL}}(k_{\rm{NL}}),
   \label{eq : Poisson eq}
\end{align}
where $k_{\rm{NL}}$ is wave number in non-linear regime and $x_{\rm{NL}}\equiv k_{\rm{NL}}\eta$.
To evaluate the $\delta_{\rm{m,NL}}(k_{\rm{NL}})$, we adopt the result of N-body simulation in Refs.~\cite{Hamilton:1991es,Peacock:1993xg}.
According to Refs.~\cite{Hamilton:1991es,Peacock:1993xg}, $\delta_{\rm{m,NL}}(k_{\rm{NL}})$ can be related to the value of $\delta_{\rm{m,L}}(k_{\rm{L}})$, which is the linearly extrapolated density perturbation, by
\begin{align}
   \delta^2_{\rm{m,NL}}(k_{\rm{NL}})
   =
   f_{\rm{NL}}\left[\delta^2_{\rm{m,L}}(k_{\rm{L}})\right],
   \label{eq : relation between NL and L}
\end{align}
where function $f_{\rm{NL}}(x)$ has fitting formula given by
\begin{align}
   f_{\rm{NL}}(x)
   =x
   \left[
   \frac{1+0.4x+0.498x^4}{1+0.00365x^3}
   \right]^{1/2},
   \label{eq : fNL}
\end{align}
and the relation between $k_{\rm{NL}}$ and $k_{\rm{L}}$ is given by
\begin{align}
   k_{\rm{L}}
   =
   \left[1+\delta^2_{\rm{m,NL}}(k_{\rm{NL}})\right]^{-1/3}
   k_{\rm{NL}}.
   \label{eq : relation between NL and L2}
\end{align}
By substituting Poisson equation of $\delta_{\rm{L}}$ to Eq.~\eqref{eq : relation between NL and L}, we obtain
\begin{align}
   \delta^2_{\rm{m,NL}}(k_{\rm{NL}})
   =
   f_{\rm{NL}}\left[
   \frac{4}{25}C^2A_s\mathcal{H}^{-4}k_{\rm{L}}^4\mathcal{T}(x_{\rm{eq,1}})^2
   \right].
   \label{eq : relation between NL and L3}
\end{align}
By substituting Eqs.~\eqref{eq : fNL} and~\eqref{eq : relation between NL and L3} to Eq.~\eqref{eq : relation between NL and L2}, we obtain the relation between $k_{\rm{NL}}$ and $k_{\rm{L}}$, denoting as $k_{\rm{NL}}=F_{\rm{NL}}(k_{\rm{L}})$.
By substituting $k_{\rm{L}}=F_{\rm{NL}}^{-1}(k_{\rm{NL}})$, Eqs.~\eqref{transfar function at eRD end} and~\eqref{eq : relation between NL and L3} into Eq.~\eqref{eq : Poisson eq}, we obtain $\mathcal{T}(x_{\rm{NL}})$. 
From now on, we write the solution $\mathcal{T}(x_{\rm{NL,dec}})$ by $\mathcal{T}(x_{\rm{NL,eMD}})=S_{\rm{NL}}\mathcal{T}(x_{\rm{eq,1}})$, where $x_{\rm{NL,eMD}}$ is the value of $x$ right before the L-ball decay. 
From now on, we omit the subscript ``$\rm{NL}$''.

Thirdly, as for era (iii), $\mathcal{T}$ decays proportionally to the matter density perturbations at first and the time evolution of $\mathcal{T}$ is given by
\begin{align}
   \frac{\mathcal{T}(x)}{\mathcal{T}(x_{\rm{eMD}})}
  \simeq
  \frac{M_{\rm{Q}}(t)}{M_{\rm{Q,init}}}\simeq \left(1-\frac{t}{t_{\rm{decay}}}\right)^{3/5},
  \label{eq : transfer func at era iii}
\end{align}
where $M_{\rm{Q}}(t)$ is the mass of each L-ball and $M_{\rm{Q,init}}$ is the L-ball mass at formation, given by Eq.~\eqref{eq : Q-ball property1}. 
The above equation assumes that gravitational potential is determined by matter density perturbation, which requires
\begin{align}
   3a^2|\ddot{\mathcal{T}}|\ll k^2\mathcal{T},
   \label{requirement that Phi determined by deltam}
\end{align}
as a necessary condition.
We assume that $\mathcal{T}$ begins to be determined by radiation density perturbation once Eq.~\eqref{requirement that Phi determined by deltam} is violated.
We denote this period by $\eta=\eta_{\rm{dcpl}}$, which satisfies
\begin{align}
   k\eta_{\rm{decay}}-k\eta_{\rm{dcpl}}=\frac{9\sqrt{2}}{5}.
   \label{eq : eta dcpl}
\end{align}
Substituting Eq.~\eqref{eq : eta dcpl} into Eq.~\eqref{eq : transfer func at era iii}, we obtain the lower bound of $\mathcal{T}$ at the onset of RD, denoted as $\mathcal{T}_{\rm{low}}$, given by 
\begin{align}
    \mathcal{T}_{\rm{low}}
    \simeq
    \left(
        \frac{9\sqrt{2}}{5k\eta_{\rm{decay}}}
    \right)^{3/5}
    S_{\rm{NL}}\mathcal{T}(x_{\rm{eq,1}})
    \equiv
    S_{\rm{decay}}S_{\rm{NL}}\mathcal{T}(x_{\rm{eq,1}})
    \ .
    \label{Phi low}
\end{align}
Note that this $\mathcal{T}_{\rm{low}}$ gives a lower bound of $\mathcal{T}$ at the onset of RD, because $\mathcal{T}$ can decouple from matter density perturbation even before the necessary condition~\eqref{requirement that Phi determined by deltam} is violated. 

Finally, we discuss era (iv).
In this period, $\mathcal{T}$ is determined by radiation density perturbations, which oscillate due to pressure on subhorizon scales. Setting $w = 1/3$, the equation of $\mathcal{T}$ is written as
\begin{align}
    \mathcal{T}'' +4\mathcal{H}\mathcal{T}'+\frac{k^2}{3}\mathcal{T}
    =
    0
    \ .
\end{align}
The solution with initial condition $\mathcal{T}(x=x_{\rm{decay}})= S_{\rm{decay}}S_{\rm{NL}}\mathcal{T}(x_{\rm{eq,1}}),\mathcal{T}'(x=x_{\rm{decay}})\simeq 0$ is given by~\cite{Kohri:2018awv,Inomata:2020tkl}
\begin{align}
    \mathcal{T}(x>x_{\rm{decay}})
    =
    S_{\rm{decay}}S_{\rm{NL}}\mathcal{T}(x_{\rm{eq,1}})(A\mathcal{J}(x)+B\mathcal{Y}(x))
    \ .
    \label{eq : transfer func during RD}
\end{align}
Here, $\mathcal{J}(x), \mathcal{Y}(x)$ are defined using first and second spherical Bessel functions, $j_1(x)$ and $y_1(x)$, as
\begin{align}
    \mathcal{J}(x)
    =
    \frac{3\sqrt{3}j_1\left(\frac{x-x_{\rm{decay}}}{\sqrt{3}}\right)}{x-x_{\rm{decay}}/2}
    \ ,
    \nonumber \\
    \mathcal{Y}(x)
    =
    \frac{3\sqrt{3}y_1\left(\frac{x-x_{\rm{decay}}}{\sqrt{3}}\right)}{x-x_{\rm{decay}}/2}
    \ ,
\end{align}
and coefficients $A$, $B$ are defined by
\begin{align}
    A
    &=
    \frac{1}{\mathcal{J}(x_{\rm{decay}})-\frac{\mathcal{Y}}{\mathcal{Y}'(x_{\rm{decay}})}\mathcal{J}'(x_{\rm{decay}})}
    \ ,
    \nonumber \\
    B 
    &=
    -\frac{\mathcal{J}'(x_{\rm{decay}})}{\mathcal{Y}'(x_{\rm{decay}})}A
    \ .
\end{align}
By substituting Eqs.~\eqref{eq : I}, \eqref{eq : f}, and~\eqref{eq : transfer func during RD} into Eq.~\eqref{eq : relation between GW spectrum and curvature spectrum}, we finally obtain the gravitational wave spectrum. The approximate analytic result is given by Eq.~\eqref{eq : approximate analytic formula of GW spectrum}. 

\small
\bibliographystyle{JHEP}
\bibliography{Ref}

\end{document}